\documentclass[12pt]{article}

\usepackage[usenames]{color}
\usepackage{ifthen}
\usepackage[latin1]{inputenc}
\usepackage[authoryear]{natbib}
\usepackage{graphicx}
\usepackage{latexsym}
\usepackage{amssymb}
\usepackage{amsmath}
\usepackage{amsfonts,amsthm,enumerate,array}
\usepackage{bm,multicol}
\usepackage{caption}
\usepackage[hidelinks,colorlinks=true,linkcolor=blue,citecolor=blue, urlcolor=blue]{hyperref}
\usepackage{chngcntr} 
\usepackage{mathrsfs} 
\usepackage{adjustbox}
\usepackage{multirow}
\usepackage{enumitem}
\usepackage{algorithm2e}
\usepackage{bbm}
\usepackage{setspace}
\usepackage{rotating}
\usepackage{mathabx}
\usepackage{ulem}
\usepackage{authblk}

\makeatletter 
\newcommand{\labitem}[2]{%
\def\@itemlabel{\textbf{#1.}}
\item
\def\@currentlabel{#1}\label{#2}}
\makeatother

\makeatletter 
\newcommand{\labitemc}[2]{%
\def\@itemlabel{\textbf{#1}}
\item
\def\@currentlabel{#1}\label{#2}}
\makeatother

\newboolean{tab}
\setboolean{tab}{false}

\makeatletter\@addtoreset{equation}{section}\makeatother

\def\logit{\text{logit}}
\def \tr{\text{tr}}
\def \bmu{\bm\mu}
\def \bSigma{\bm\Sigma}
\def \R{\mathbb{R}}
\def \E{\mathbb{E}}

\newcommand{\bean}{\begin{eqnarray*}}
\newcommand{\eean}{\end{eqnarray*}}
\newcommand{\bea}{\begin{eqnarray}}
\newcommand{\eea}{\end{eqnarray}}
\newcommand{\be}{\begin{eqnarray}}
\newcommand{\ee}{\end{eqnarray}}
\newcommand{\beq}{\begin{equation}}
\newcommand{\eeq}{\end{equation}}

\renewcommand{\hat}{\widehat}
\renewcommand{\tilde}{\widetilde}
\renewcommand{\check}{\widecheck}

\advance\topmargin by -0.85truein
\advance\textheight by 1.25truein

\tiny\normalsize
\date{}
\begin{document}
\thispagestyle{empty}
\title{Bayesian Unit-level Modeling of Categorical Survey Data with a Longitudinal Design}

\author[1,$\dagger$]{Daniel Vedensky}
\author[2]{Paul A. Parker}
\author[1,3]{Scott H. Holan}

\affil[1]{{Department of Statistics}, {University of Missouri}}
\affil[2]{{Department of Statistics}, {University of California, Santa Cruz}}
\affil[3]{{Research and Methodology Directorate, U.S. Census Bureau}}

\maketitle 
\begingroup
\renewcommand{\thefootnote}{} 
\footnotetext{$^\dagger$To whom correspondence should be addressed. \url{dvedensky@mail.missouri.edu}}
\endgroup

\abstract{
Categorical response data are ubiquitous in complex survey applications, yet few methods model the dependence across different outcome categories when the response is ordinal.
Likewise, few methods exist for the common combination of a longitudinal design and categorical data.
By modeling individual survey responses at the unit-level, it is possible to capture both ordering information in ordinal responses and any longitudinal correlation.
However, accounting for a complex survey design becomes more challenging in the unit-level setting.
We propose a Bayesian hierarchical, unit-level, model-based approach for categorical data that is able to capture ordering among response categories, can incorporate longitudinal dependence, and accounts for the survey design.
To handle computational scalability, we develop efficient Gibbs samplers with appropriate data augmentation as well as variational Bayes algorithms. 
Using public-use microdata from the Household Pulse Survey, we provide an analysis of an ordinal response that asks about the frequency of anxiety symptoms at the beginning of the COVID-19 pandemic. 
We compare both design-based and model-based estimators and demonstrate superior performance for the proposed approaches.\\

}

\doublespace
\newpage

\section{Introduction}\label{sec:intro}
In the context of survey data applications, modeling individual records (i.e., ``unit-level modeling") rather than area-level aggregates is often necessary.
For ordinal-valued responses, modeling the individual survey responses allows us to capture dependence among the response categories, which is not possible in area-level models that model group averages.
Likewise, for longitudinal studies, modeling at the unit-level is necessary in order to link individual responses and capture the serial correlation across a given respondent's survey responses.
In small area estimation applications, unit-level models have the additional benefit of allowing for arbitrary aggregation of data and may increase the precision of estimates \citep{hid16}.
Yet, despite the ubiquity of non-Gaussian survey data collected according to a longitudinal design, there is scarce unit-level methodology for such settings.
Notably, there are few existing methods for the combination of ordinal and longitudinal data that are able to incorporate spatio-temporal dependence and account for complex survey designs.

In general, more attention has been paid to binary and nominal responses  in the 
complex survey literature \citep{Skinner_2018}.
{\color{black}In contrast, in the spatial literature for complex surveys}, categorical data are typically modeled at the area level, which requires calculating weighted proportions first. 
When the data follow a natural ordering, it is common to ignore this structure and instead model the ordinal proportions as though they were nominal.
For example, \citet{bauder_2021} model multinomial probabilities as multivariate normal {\color{black}after employing} a logit transformation.
Although it is possible to induce an ordinal structure by modifying the covariance matrix \citep{agresti_2010}, this does not capture the dependence a unit-level approach would.

Meanwhile, most existing unit-level methodology for categorical survey data does not fully take into account the survey design or applies only to cross-sectional surveys.
For example,
 \citet{Machini_2022} present an ordinal model for spatial data, but disregard the survey design,
\citet{beltran_2024} model ordinal survey data with spatial dependence, but rely on the restrictive assumption that all design information is available in the form of covariates.
Methods that do account for a longitudinal survey design generally assume Gaussian, binary, or occasionally nominal data.
One exception is the longitudinal model of \citet{kunihama_2019} that is able to handle multivariate responses and incorporates the weights into a Bayesian nonparametric (BNP) sampling algorithm. 
However, the response types considered are continuous, binary, or nominal, but not ordinal.
\citet{sutradhar_2000} consider ordinal, longitudinal, complex design, but use GEE which does not easily extend to the spatial setting or allow for straightforward uncertainty quantification.

In this paper, we propose an ordinal model within a unit-level modeling framework that incorporates survey weights, spatial, and temporal dependence.
We do so by extending the work of \citet{parker_holan_janicki_2022} and \citet{vedensky_2023}.
The former models cross-sectional binary and nominal complex survey data at the unit level while the latter builds on this approach to cover longitudinal data with Gaussian or binary responses. 
By employing a particular stick-breaking representation of a multinomial likelihood \citep{linderman2015}---also known as a continuation ratio factorization \citep{fienberg_2007}---along with P\'olya-Gamma data augmentation \citep{polson_scott_2013}, we obtain conjugate sampling algorithms, with Bayesian, ordinal logistic regression as a special case.  
When data are especially high-dimensional, MCMC algorithms may prove insufficient, and we introduce Variational Bayes (VB) algorithms \citep{blei_2017} as an alternative to Gibbs sampling for such applications.

Recent work has made use of this combination of the continuation ratio factorization and P\'olya-Gamma data augmentation. 
\citet{kang_crosssectional} develop a BNP model for cross-sectional data
and \citet{kang_longitudinal} extend this to functional, longitudinal data, modeling continuous curves. 
In the psychological literature, \citet{jimenez2023} cover the special case of an item-response model with ordinal values.
 However, modifying these approaches to handle complex survey data with a spatial component is nontrivial, and to our knowledge, the link to Bayesian logistic regression has not been pointed out elsewhere.

To illustrate the efficacy of our proposed methodology, we apply our models to data from the Household Pulse Survey (HPS), a high-dimensional complex survey that follows a longitudinal design.
This includes an empirical simulation study and a full data analysis.

This article proceeds as follows. 
In Section~\ref{sec:uw_ord} we describe our general approach to modeling ordinal data.
Section~\ref{sec:survey} describes how the ordinal and nominal models may be extended for use with survey data.
In Section~\ref{sec:vb} variational Bayes algorithms for each of the models are introduced.
We then present empirical results in Section~\ref{sec:results} and conclude with a discussion in Section~\ref{sec:discuss}.

\section{Bayesian Ordinal Logistic Regression}
\label{sec:uw_ord}

Let $\mathcal{U} = \{1,\ldots, N\}$ index a population, from which we observe a sample $\mathcal{S}\subseteq \mathcal{U}$ of size $|\mathcal{S}|=n$. 
Associated with each element of the sample are a set of covariates $\bm x_i = (x_{i1},\ldots, x_{iq})$ as well as an ordinal-valued response $y_i \in \{1, \ldots , K\}$, with $K>2$ possible categories.
We also define the one-hot encoded vector representation of the response, $\tilde{\bm{y}}_i = (\mathbbm{1}(y_i = 1), \mathbbm{1}(y_i=2), \ldots, \mathbbm{1}(y_i=K))$, where $\mathbbm{1}(\cdot)$ denotes the indicator function.
The response vectors for the entire sample may be collated into the $nK$-dimensional vector $\tilde{\bm y} = (\tilde{\bm y}_1, \tilde{\bm y}_2, \ldots, \tilde{\bm y}_n).$

It is common to model such data with a \textit{cumulative} model \citep{mccullagh_1980}, where latent, ordered cutpoints $0\leq\gamma_1\leq\cdots \leq \gamma_{K-1}$ are introduced and it is assumed
\[P(y_i \leq k) = F(\gamma_k - \bm x_i'\bm\beta),\]
with $F$ being any strictly monotonic distribution function and ${\bm\beta}$ a set of regression coefficients.
This implies the probability of observing a given category to be
\[P(y_i = k) = F(\gamma_k - \bm x_i'\bm\beta) - F(\gamma_{k-1} - \bm x_i'\bm\beta).\]

In the Bayesian setting, \citet{albert_chib1993} propose a popular data augmentation algorithm for cumulative, ordinal probit regression. 
They take $F$ to be the standard normal cdf $\Phi$ and add Gaussian latent variables to produce an easily implementable Gibbs sampler.
This is the most widely used Bayesian formulation for ordinal models \citep{agresti_2010}, 
and has been adapted to spatial models, as well \citep{higgs_2010, schliep_2015, carter_2024}.

\citet{tutz_1990, tutz_1991} specify an alternative latent variable formulation for a \textit{sequential} ordinal model, which has received relatively less attention in the literature.
However, a number of authors have noted benefits, such as more flexible modeling of response curves and the possibility for category-specific covariates \citep{tutz_2005, burkner_2019, boes_2006, kang_crosssectional}. 
In contrast to the cumulative approach, the sequential approach assumes no ordering for $\bm\gamma=(\gamma_1,\ldots, \gamma_{K-1})$, and that the conditional probability for observing a given category $k<K$ is
\begin{equation}
  \label{eq:tutz1}
  P(y_i = k| \bm\beta, \bm\gamma) = F(\gamma_k-\bm x_i'\beta)\prod_{j<k}(1-F(\gamma_j-\bm x_i'\bm\beta))
\end{equation}
and 
\begin{equation}
  \label{eq:tutz2}
P(y_i = K| \bm\beta, \bm\gamma) = \prod_{j=1}^{K-1}(1-F(\gamma_j-\bm x_i'\bm\beta))
\end{equation}
for the last category.
In this way, the probability of observing a particular category may be viewed as a sequential process, where category $k$ can only be attained if all previous categories, $j<k$ are not observed.
The final category, $K$, is observed only if all other categories fail to be attained.

This formulation implies an independent binomial model for each response category, conditional on $\bm\beta$ and $\bm\gamma$,
via the continuation-ratio parametrization of \citet{fienberg_1980}, which is also referred to as a ``stick-breaking" representation \citep{linderman2015}.
Specifically, a multinomial pmf parameterized by counts $n$ and a vector of probabilities $\bm \pi_i=(\pi_{i1},\ldots,\pi_{iK})$, can be expressed as the product
\begin{equation}
  \label{eq:mult_sb}
  \text{Multinomial}({\tilde{\bm y}}_i|n_i,\bm \pi_i) = \prod_{k=1}^{K-1}\text{Binomial}(\tilde y_{ik}|n_{ik},\tilde \pi_{ik}),
\end{equation}
where $\pi_{ik} = \tilde \pi_{ik}\prod_{j<k}(1-\tilde\pi_{ij})$ and $n_{ik}=n_i-\sum_{j<k} y_j$ for $k=1,\ldots, K-1$. That is, if we let $\pi_{ik} = P(y_i = k | \bm\beta,\bm\gamma)$ and $\tilde \pi_{ik} = F(\gamma_k-\bm x_i\bm\beta)$, with $F$ the inverse logit function, we have that (\ref{eq:tutz1}) and (\ref{eq:tutz2}) define a multinomial likelihood over the ordinal categories.

Placing Gaussian priors on $\bm\beta$ and $\bm\gamma$, we may then specify the full ordinal logistic regression model hierarchically as a Bayesian GLM
\begin{align} 
\nonumber
  p(\bm{\tilde y}|\bm\beta) &= \prod_{i\in\mathcal{S}}\prod_{k=1}^{K-1} \text{Binomial}(\tilde y_{ik}|n_{ik}, \tilde{\pi}_{ik})\\  \nonumber
  \logit(\tilde \pi_{ik}) &= \gamma_k - \bm x_i'\bm\beta \label{eq:ord_mod}\\ 
  \bm\beta &\sim \text{N}_q(\bm 0, \sigma^2_\beta \text{I}_q)\\\nonumber
  \bm\gamma &\sim \text{N}_{K-1}(\bm 0, \sigma^2_\gamma I_{K-1})\\\nonumber
   \sigma_\beta, \sigma_\gamma &> 0.\nonumber
\end{align}

For nominal responses, where there is no ordering of response categories, model (\ref{eq:ord_mod}) simplifies to estimating $\logit(\tilde{\pi}_i) = \bm x_i\bm\beta_k$ with category-specific coefficients, $\bm\beta_k$ and with no need  for cutpoints $\gamma_k$ \citep{parker_holan_janicki_2022}
\begin{align} 
\nonumber
  p(\bm{\tilde y}|\bm\beta) &= \prod_{i\in\mathcal{S}}\prod_{k=1}^{K-1} \text{Binomial}(\tilde y_{ik}|n_{ik}, \tilde{\pi}_{ik})\\  \nonumber
  \logit(\tilde \pi_{ik}) &= \bm x_i'\bm\beta_k \label{eq:nom_mod}\\ 
  \bm\beta_k &\sim \text{N}_q(\bm 0, \sigma^2_\beta \text{I}_q)\\\nonumber
   \sigma_\beta &> 0.\nonumber
\end{align}

Because both (\ref{eq:ord_mod}) and (\ref{eq:nom_mod}) have logistic likelihoods, it is possible to introduce latent variables that follow a P\'olya-Gamma (PG) distribution \citep{polson_scott_2013} to obtain computationally efficient Gibbs samplers.
In the case of (\ref{eq:ord_mod}) this yields a conjugate sampler for logistic ordinal regression.

A random variable $X$ is said to be distributed PG$(b, c)$ with shape parameter $b>0$ and scale parameter $c\in \R$ if it is equal in distribution to 
\[\frac{1}{2\pi^2}\sum_{k=1}^\infty \frac{g_k}{(k-1/2)^2+c^2/(4\pi^2)},\]
where $g_k\overset{ind.}{\sim} \text{Gamma}(b, 1)$.
\citet{polson_scott_2013} also show that
\begin{equation}
\E[X] = \frac{b}{2c}\tanh(c/2) \label{eq:PG_expectation}
\end{equation}
and derive the integral identity,
\begin{equation}\frac{e^{a\lambda}}{(1+e^{\lambda})^b} = 2^{-b}e^{\kappa\lambda}\int_0^\infty e^{-\omega\lambda^2/2}p(\omega)d\omega, \label{eq:integral_identity}
\end{equation}
where $\kappa=a-b/2$ and $\omega\sim PG(b,0)$. 
In the right-hand side of (\ref{eq:integral_identity}), after conditioning on $\omega$, we have a Gaussian density in terms of $\lambda$. 
If we condition on $\lambda$, then we have a PG density in $\omega$.
Therefore, sampling the model only requires alternating between Gaussian and PG random draws.

Notation for the full sampling algorithm is greatly simplified if, following \citet{albert_2001}'s algorithm for probit regression, we augment the covariate vector for each response such that
if $y_i = k$, then 
$\bm{\tilde{x}}_{ij} = (0,\ldots,  1,\ldots, 0, -\bm x_i')$,  for $j=1,\ldots, \min(k,K-1)$, where the $1$ is in the $j$th position. 
 Further, let $\bm{\tilde X}_i = (\tilde{\bm x}_{i1}, \ldots, \tilde{\bm x}_{ij})'$ be the new covariate matrix for respondent $i$ and 
\begin{equation}
  \label{eq:augmented_matrix}
  \bm{\tilde X} = \begin{pmatrix} \tilde{\bm X}_{1}\\
                             \vdots\\
                             \tilde{\bm X}_{n}
\end{pmatrix}
\end{equation}
the augmented design matrix for the entire sample.
Letting $\bm\delta = (\bm\gamma', \bm\beta')',$ we then have
\[ \logit^{-1}(\gamma_k - \bm x_i'\bm\beta)  = \logit^{-1}(\tilde{\bm x}_{ik}' \bm\delta) = \frac{\exp(\tilde{\bm x}_i'\bm\delta)}{1+\exp(\tilde{\bm x}_i'\bm\delta)},\]
so that the integral identity (\ref{eq:integral_identity}) allows us to write
the multinomial data model in (\ref{eq:ord_mod}) as 
\[\prod_{i\in\mathcal{S}}\prod_{k=1}^{K-1}\frac{1}{2}e^{\kappa_{ik}\tilde{\bm x}_i'\bm\delta}\int_0^\infty e^{-\omega_{ik}(\tilde{\bm x}_i'\bm\delta)^2/2}p(\omega_{ik})d\omega_{ik},\]
with $\kappa_{ik}=y_{ik}-1/2.$

Assuming the prior distribution $\bm\delta\sim N(\bm d, \bm D)$, a Gibbs sampler for the model requires iterating over only two steps
\begin{enumerate}
\item  $\omega_{ik}|\bm\delta \sim \text{PG}(1, \bm{\tilde x_{ik}}'\bm\delta)$  (for $i=1,\ldots,n$ and $k=1,\ldots,k_i$)
\item  $\bm \delta|\bm \Omega \sim \text{N}(\bm\mu_\delta, \bm\Sigma_\delta)$, 
\end{enumerate}
where $\bm\Sigma_\delta = (\bm D^{-1}+\bm{\tilde{X}'\Omega \tilde{X}})^{-1}$ and $\bm\mu_\delta = \bm\Sigma_\delta(\bm{\tilde{X}}'\tilde{\bm{\kappa}} + \bm D^{-1}\bm d)$ with $\bm\omega = (\omega_{11}, \omega_{12}, \ldots, \omega_{nk_n})$ and $\bm\Omega=\text{diag}(\bm\omega).$
The vector $\tilde{\bm{\kappa}}$ is constructed by the same process as (\ref{eq:augmented_matrix}), where $\tilde{\bm{\kappa}}_i = (\kappa_{i1}, \ldots, \kappa_{ij})'$ and $\tilde{\bm\kappa}=(\tilde\kappa_1', \ldots, \tilde\kappa_n')',$

\section{Categorical Models for Complex Survey Data}\label{sec:survey}

In order to apply the above models to unit-level survey data, it is important to consider a number of additional factors.
First, complex surveys often collect data according to a design that may not be known to the data modeler \citep{gelman2007}.
Even if the design is known, factors such as non-response, attrition (in the case of longitudinal surveys) \citep{thompson2015}, and informative sampling \citep{Parker_2023} may cause bias.
These complications must be accounted for through proper use of survey weights.
Large-scale surveys are also likely to exhibit spatial dependence and it is common to produce predictions for spatial (or demographic) domains with low sample sizes (i.e., small area estimation).
Additionally, many surveys are conducted repeatedly over time, or employ a longitudinal structure, where the same people or households are re-interviewed, leading to temporal correlation. 
The Bayesian hierarchical modeling framework allows us to address each of these types of dependence by extending the data and process model components of (\ref{eq:ord_mod}).

\subsection{Complex survey design}

In area-level models, the survey design is incorporated into the direct estimates being modeled. 
For instance, suppose our population $\mathcal{U}$ can be partitioned into $m$ disjoint areas $\mathcal{U}_j$ such that $\mathcal{U} = \bigcup_{j=1}^m \mathcal{U}_j$.
In a complex survey, a sample is taken according to known selection probabilities $\pi_{ij} = P(i\in\mathcal{S}_j)$ and weights are defined as $w_{ij}=1/\pi_{ij}$.
To fit an area-level model, a direct estimator such as the Horvitz-Thompson estimator for the mean \citep{Horvitz1952} 
\begin{equation}
\hat{\bar y}_j = \frac{1}{|\mathcal{U}_j|}\sum_{i\in\mathcal{S}_j} w_{ij}y_{ij}
\label{eq:horvitz_thompson}
\end{equation}
is calculated for each area and treated as the response variable.
The inclusion of survey weights in (\ref{eq:horvitz_thompson}) reflects the survey design, hence there is no bias due to an unrepresentative sample.

Unit-level models, on the other hand, are fit to individual survey responses, so incorporating survey weights becomes less straightforward.
\citet{Parker_2023} provide an overview of remedies for informative sampling/complex design in this setting.
In particular, the sequential ordinal regression model readily fits into the pseudo-likelihood framework \citep{binder_1983, skinner_1989}, in which the likelihood contribution of each unit is exponentiated by its survey weight, 
\[\prod_{i\in\mathcal{S}}f(y_{i}|\bm\theta)^{w_{i}}.\]
\citet{savitsky_toth_2016} show that in the Bayesian case, a pseudo-likelihood combined with a prior density $p(\bm \theta)$, leads to a valid pseudo-posterior
\[p(\bm\theta|\bm y, \bm{\tilde w}) \propto \prod_{i\in\mathcal{S}} f(y_i|\bm\theta)^{\tilde{w}_i}p(\bm\theta)\]
provided that the original survey weights are rescaled to $\tilde{w}_i = nw_i/\sum_{i\in\mathcal S} w_i$ in order to ensure $\sum\tilde{w}_i=n$.

In the case of (\ref{eq:ord_mod}), introducing a pseudo-likelihood means that the binomial pmf in the data model, which is parameterized by log odds $F(\gamma_k-\bm x_i'\bm\beta)$, becomes proportional to
\begin{align*}
  [(F(\gamma_k - \bm x_i'\bm\beta))^{y_{ik}}(1-F(\gamma_k - \bm x_i'\bm\beta))^{1-y_{ik}}]^{\tilde{w}_i} &=
  \frac{(\exp(\gamma_k-\bm x_i'\bm\beta))^{y_{ik}\tilde{w}_i}}{(1+\exp(\gamma_k-\bm x_i'\bm\beta))^{n_{ik}\tilde{w}_i}}
,
\end{align*}
which is the integral identity (\ref{eq:integral_identity}) with $a=y_i\tilde w_i$ and $b=n_i\tilde w_i$.
Consequently, sampling again proceeds by drawing iteratively from PG and Gaussian distributions, but with the PG shape parameter now containing a term for the survey weights so that $\omega_{ik}|\bm\delta \sim \text{PG}(\tilde{w}_in_{ik},\tilde{\bm x}'_{ik}\bm\delta)$ in the first step.

\subsection{Spatio-temporal dependence}
\label{subsec:spatio-temporal}
Survey data often contain spatial and temporal information.
If the spatial domain of interest is especially fine-grained, then many areas will have small sample sizes and we would like to leverage spatial dependence to ``borrow strength."
This can be achieved by including an area-level random effect within a unit-level model.
\citet{parker_holan_janicki_2022} propose Bayesian hierarchical models for cross-sectional data with binary, count, and categorical responses.
They introduce an $m-$dimensional vector of areal random effects $\bm\eta\sim N_m(\bm 0, \sigma^2_\eta I_m)$, which retain conjugacy with the P\'olya-Gamma data augmentation scheme introduced above.
To induce a more explicit spatial dependence structure, they take $\bm\psi_i$ to be a set of spatial basis functions, constructed as a subset of the eigenvectors of the adjacency matrix for all states. %
In our model, we retain only the $m$ columns corresponding to positive eigenvalues (and hence, positive spatial dependence), but for higher-resolution geographies further truncation can be performed to achieve greater dimension reduction.
Thus, the $n\times m$ matrix of basis functions is defined as 
\[\bm\Psi=\begin{pmatrix}\bm\psi_1'\\ \vdots\\ \bm\psi_n'\end{pmatrix},\] 
with $\bm\psi_i$ the basis function evaluated at the area containing unit $i$.
This construction may be viewed as a special case of the Moran's I basis functions \citep{hughes2013, bradley_2015}.
Incorporating such a random effect into (\ref{eq:ord_mod}) and (\ref{eq:nom_mod}) leads us to set the linear predictor to be $\logit(\tilde \pi_{ik}) = \gamma_k - (\bm x_i \bm\beta + \bm\psi_i \bm\eta).$

Many complex surveys  follow a longitudinal design wherein the same respondents (or some subset of them) are re-interviewed over time. %
Two types of dependence arise in this setting:
within-unit responses are correlated, as are the aggregate-level trends.
For longitudinal designs where the survey is repeated at regular intervals, \citet{vedensky_2023} present models for Gaussian, binary and count data that handle both types of correlation. 
For discrete responses, the within-unit correlation is captured by constructing a categorical covariate that indexes the value (or absence) of the respondent's previous response.
The aggregate-level dependence is captured by adding an AR(1) structure onto the areal random effect so that the process model follows
\begin{align}
\label{eq:lon_reff1}
\bm \eta_t |\bm\eta_{t-1},\phi,\sigma^2_{\eta} &\sim \text{N}_m(\phi\bm\eta_{t-1}, \sigma^2_{\eta} I_m), \qquad \text{ for } t=2,\ldots,T\\
\bm \eta_1 |\sigma^2_{\eta} &\sim \text{N}_m(\bm 0, \sigma^2_{\eta} I_m),
\label{eq:lon_reff2}
\end{align}
where $\phi \sim \text{Unif}(-1,1)$ functions as an autoregressive parameter, $\sigma^2_\eta,\sigma^2_{\eta_1} \overset{ind.}{\sim} IG(a,b)$ are the prior random effect variances, 
and $a,b, \sigma_\beta > 0$ are fixed hyperparameters.

We note that a survey may also follow a repeated, cross-sectional design, where the entire sample is refreshed at each time point.
In such a survey, there is no within-respondent correlation and the random effects structure (\ref{eq:lon_reff1})--(\ref{eq:lon_reff2}) alone is sufficient for capturing the design.

Combining the above modifications, the full spatio-temporal, ordinal model can be specified as
  \begin{align*}
  \bm y | \bm \beta, \bm\eta, \bm\gamma &\propto \prod_{t=1}^T \prod_{i\in\mathcal{S}_{t}} \prod_{k=1}^{K-1} \text{Binom}(y_{i{t}k}|n_{i{t}k},\tilde \pi_{i{t}k})^{\tilde w_{i{t}}}\\
   \text{logit}(\tilde \pi_{i{t}k}) &= {\bm u_{itk}'\bm\gamma} - \bm x_i'\bm\beta - \bm\psi_i \bm\eta_{ t}\\
   \bm\gamma &\sim \text{N}_{g}(0, \sigma^2_\gamma I_{g}) \\
{\bm\eta_t|\bm\eta_{t-1},\phi, \sigma^2_\eta\phantom{.}} &\sim \text{N}_m(\phi\bm \eta_{t-1}, \sigma^2_\eta I_m),\hspace{2mm} t=2,\ldots, T\\
{ \bm\eta_1|\sigma^2_{\eta}\phantom{.}} &\sim {\text{N}_m(\bm 0, \sigma^2_{\eta} I_m)}\\
   \bm\beta &\sim \text{N}_q(\bm 0, \sigma^2_\beta I_q)\\
   {\phi }\phantom{.}&\sim \text{Unif}(-1,1)\\
   \sigma^2_\eta &\sim \text{IG}(a,b),
  \end{align*}
  where $n_{itk}=n_{it}-\sum_{j<k} y_{itj}$ and $\widetilde \pi_{itk}=\pi_{itk}/(1-\sum_{j<k} \pi_{itj})$.
  One additional feature of the model is that the cutpoints are made to vary by time and according to a subject's previous response.
  Therefore, there are $g= (K-1) + (T - 1)*(K+1)*(K-1)$ cutpoints because subjects in the first time point have no previous response, whereas for the remaining $T-1$ weeks, there are $K+1$ possible values for the previous response.
  Each of these scenarios requires $K-1$ cutpoints.
  The indicator vector $\bm u_{itk}$ picks out the appropriate cutpoint for a response.%

For unordered data, a similar nominal model can be fit as $K-1$ independent binomial models, but with no cutpoints and with category-specific fixed and random effects
\begin{align*}\label{eq:mult} 
  \begin{split}
\bm y| \bm\beta, \bm\eta &\propto \prod_{t=1}^T\prod_{i\in \mathcal{S}_t}\prod_{k=1}^{K-1} \text{Binomial}(y_{itk} | n_{itk}, \tilde \pi_{itk})^{\widetilde{w}_{it}}\\
\text{logit}(\widetilde \pi_{itk}) &= \bm x_i'\bm\beta_k+\bm\psi_i'\bm\eta_{tk}\\
\bm \eta_{tk} |\bm\eta_{(t-1)k},\phi_k,\sigma^2_{\eta k} &\sim \text{N}_m(\phi_k\bm\eta_{(t-1)k}, \sigma^2_{\eta k} I_m), \qquad  t=2,\ldots,T; \phantom{--} k = 1,\ldots K-1\\
\bm \eta_{1k} |\sigma^2_{\eta_1k} &\sim \text{N}_m(\bm 0, \sigma^2_{\eta_1k} I_m) \qquad k=1,\ldots,K-1\\
\bm \beta_k &\sim \text{N}_q(\bm 0, \sigma^2_\beta I_q) \qquad  k=1,\ldots,K-1\\
\phi_k &\sim \text{Unif}(-1,1)\\
\sigma^2_{\eta_1k}, \sigma^2_{\eta k}, &\overset{ind.}{\sim} IG(a,b)\\
\sigma_\beta, a, b &> 0.
\end{split}
\end{align*}
In all of the above models, the latent processes and priors are either conditionally conjugate with the P\'olya-Gamma latent variables or produce known conditional distributions. 
The resulting Gibbs samplers are thereby efficient and straightforward to implement. 
Full details of the sampling algorithms and derivations of full conditionals are included in the Appendix.

\section{Variational Bayes}
\label{sec:vb}
Despite the conjugacy in the above models, the Gibbs samplers may be inadequate for extremely high-dimensional problems, which require sampling a large number of latent variables and random effects (e.g., large sample sizes and fine-scaled geographies).
In this case, we also propose a set of variational Bayes algorithms \citep{bishop_2006, blei_2017} as an alternative to MCMC.
These algorithms are much faster albeit at the cost of approximating the posterior distribution rather than sampling from the exact posterior.

Instead of sampling sequentially, VB methods minimize the Kullback-Leibler divergence \citep{kullback_1951}
\[\text{KL}(q(\bm\theta)\mid\mid p(\bm\theta| \bm x)) = \int_{\Theta} q(\bm\theta)\log\frac{q(\bm\theta)}{p(\bm\theta|\bm x)}d\bm\theta.\]
between the true posterior $p(\theta| x)$ and an approximation $q(\theta)$ selected from a given class of distributions $\mathcal{Q}$.
That is, an optimal approximation is selected as
\[q^*(\bm\theta) = \mathop{\arg \min}\limits_{q(\bm\theta)\in\mathcal{Q}}\text{KL}(q(\bm\theta)\mid\mid p(\bm\theta \mid \bm x)).\]
A common choice for $\mathcal{Q}$ is the mean-field variational family \citep{blei_2017} which consists of all densities $q$ that can be factored into mutually independent ``variational factors'' $q_j(\theta_j)$ for each of $J$ latent variables so that 
\[q(\bm\theta)= \prod_{j=1}^J q_j(\theta_j).\]

Since the KL divergence is not necessarily computable, the optimization problem may be reformulated as a maximization with respect to the evidence lower bound (ELBO)  \citep{blei_2017}
\[\text{ELBO}(\bm\theta) = \E[\log p(\bm\theta,\bm x)] - \E[\log q(\bm\theta)].\]
\citet{durante2019}, building on earlier work of \citet{jaakkola2000}, show that logistic models augmented by P\'olya-Gamma latent variables have a particularly simple VB algorithm for performing this maximization, and we adapt this to the ordinal case.

Returning to the notation of the two-stage Gibbs sampler at the end of Section~\ref{sec:uw_ord}, note that the Gaussian and PG densities therein have the exponential family representation
\[p(\bm \delta|\bm y, \bm \omega) \propto \exp[\bm\eta_1(\bm y)'\bm\delta + \text{vec}[\bm\eta_2(\bm \omega)]'\text{vec}(\bm\delta\bm\delta') - \alpha[\bm\eta_1(\bm y), \bm\eta_2(\bm \omega)]]\]
and
\[p(\bm \omega_{ik}|\bm y, \bm \omega) \propto \exp\{\eta_{ik}(\bm \delta)\omega_{ik} - \alpha[\eta_i(\bm \delta)]\}p(\omega_{ik}), \text{ for } i=1,\ldots, n; k=1,\ldots, k_i\]
respectively, with natural parameters $\bm\eta_1(\bm y) = \bm{\tilde{X}}'\tilde{\bm{\kappa}} + \bm D^{-1}\bm d$ 
and $\bm\eta_2(\bm\omega) = -\frac{1}{2}\bm D^{-1}+\tilde{\bm X}'\bm\Omega\tilde{\bm X}$ 
and $\eta_{ik}(\bm\omega) = -\frac{1}{2}(\tilde{\bm x}_{ik}'\bm\beta)^2.$
\citet{durante2019} show that the optimal CAVI parameter value updates at iteration $j$ are
\begin{enumerate}
    \item $\bm\lambda_1^{(j)} = \E_{q^{(j-1)(\bm{\omega})}}[\bm\eta_1(\bm y)]$%
    \item $\bm\lambda_2^{(j)} = \E_{q^{(j-1)}(\bm{\omega})}[\bm\eta_2(\bm\omega]$ %
    \item $\check\xi_{ik}^{(j)} = \E_{q^{(j)}(\bm\delta_{ik})}[\eta_{ik}(\bm\omega)] $ %
     for $i = 1, \ldots, n$ and $k=1,\ldots, k_i$,
\end{enumerate}
where the $q^{(r)}(\bm\delta_{ik})$ densities are Gaussian and the $q^{(r)}(\omega_{ik})$ densities are PG.

Calculation of the expectation in Step 2 is straightforward due to the closed-form expression provided by (\ref{eq:PG_expectation}). 
Convergence is measured by comparing the change in successive iterations of the ELBO and once this value drops below a predetermined threshold, the algorithm terminates.
Independent samples  are then drawn from the variational posterior N$(\check\bmu, \check \bSigma)$, as an approximation to the true posterior, with $\check \bmu=(-2\bm\lambda_2)^{-1}\bm\lambda_1$ and $\check \bSigma=(-2\bm\lambda_2)^{-1}$.

\citet{parker_holan_janicki_2022} adapt this approach to cross-sectional binary and nominal pseudo-likelihood models with random effects.
We further develop a cross-sectional, ordinal pseudo-likelihood sampling procedure in Algorithm \ref{alg:vb_ord_cs} and a longitudinal variant in Algorithm \ref{alg:vb_ord_long}.
The matrices $\tilde{G}$, $\tilde{X}$ and $\tilde{\Psi}$ are constructed in the same manner as (\ref{eq:augmented_matrix}), %
where if $y_{it}=k$, then we define $j_{it} = \min(k, K-1)$ and $\tilde{G}_{it} = (\bm u_{it1},\ldots, \bm u_{itj_{it}})$ and let
\begin{equation*}
\tilde{G} = \begin{pmatrix}
  \tilde{G}_{11}\\
  \tilde{G}_{12}\\
  \vdots\\
  \tilde{G}_{nT}
            \end{pmatrix}
\end{equation*}
and similarly for $\tilde{X}= (\bm x_{it1},\ldots, \bm x_{itj_{it}})'$ and $\tilde{\Psi}= (\bm \psi_{it1},\ldots, \bm \psi_{itj_{it}})'$.%

In Algorithm~\ref{alg:vb_ord_long}, calculations for the variational parameters $\tilde{\mu}_\phi$ depend on moments of the truncated normal distribution \citep{johnson_1994}.
In that algorithm, $\Phi$ denotes the standard normal cdf and $\varphi$ the pdf.
Note, too, the distinction between $(\tilde{\mu}_\phi)^2$ and $\tilde{\mu}_{\phi^2}$.
The former is the squared mean, while the latter is the variational variance parameter for $\phi$.
Subscripts on the matrices, such as $X_t$, refer to the subset of rows corresponding to respondents at time $t$.
A subscript such as, $\check{\bm \Sigma}_{\eta_{-T}}$ refers to all entries of the matrix except those corresponding to the last time point $t=T$.
\normalem
\RestyleAlgo{ruled} 
\begin{algorithm}[ht!]
\caption{Variational Bayes algorithm for the cross-sectional ordinal model\label{alg:vb_ord_cs}}
\KwData{$\tilde{\bm C}=[\widetilde{\bm G}, -\tilde{\bm X}, -\tilde{\bm \Psi}]$ %
          and $\tilde{\bm\kappa}$;}
Initialize $\check{\sigma}^2_\eta$ and $\check{\xi}_{ik}$ for $i=1,\ldots,n$ and $k=1,\ldots, k_i$;\\
\For{$j=1$ until convergence}{
  $\bm{\check\Omega} \gets \text{Diag}\left(\frac{\tilde w_{11}}{2\xi_{11}}\tanh(\check\xi_{11}/2),\ldots,\frac{\tilde w_{nk_n}}{2\xi_{nk_n}}\tanh(\check\xi_{nk_n}/2)\right);$\\
  $\check \bSigma \gets \left(\text{blockdiag}\left(\frac{1}{\sigma^2_\gamma}I_{K-1},\frac{1}{\sigma^2_\beta}I_q, {\check\sigma^2_\eta}I_m\right) +\tilde{\bm C}'\bm{\check\Omega}\tilde{\bm C}\right)^{-1};$\\
  $\check \bSigma_\eta \gets \check{\bSigma}[(g+q+1):(g+q+m),(g+q+1):(g+q+m)];$\\
  $\check\bmu \gets (\check\bmu_\gamma', \check\bmu_\beta',\check\bmu_\eta')' \gets \check\bSigma\tilde{\bm C}'\tilde{\bm\kappa};$\\
  $\check{\sigma}^2_\eta \gets (b + \frac{1}{2}(\check{\bmu}_\eta'\check{\bmu}_\eta + \text{tr}(\check\bSigma_\eta)))/(a+m/2);$\\
  \For{$i=1$ to $n$}
  {\For{$k=1,\ldots,k_i$}{$\check \xi_{ik} \gets (\tilde{\bm C}_{ik}'\check\bSigma \tilde{\bm C}_{ik} + (\tilde{\bm C}_{ik}'\check\bmu)^2)^{1/2};$\\}}
}
\end{algorithm}
\clearpage

\begin{algorithm}[ht!]
\small
\caption{Variational Bayes algorithm for the longitudinal ordinal model\label{alg:vb_ord_long}}
\KwData{$\tilde{\bm C}=[\tilde{\bm G}, -\tilde{\bm X}, -\tilde{\bm \Psi}]$ and $\tilde{\bm\kappa}$}
Initialize $\check{\bm\mu}_\gamma$, $\check{\bm{\mu}}_{\eta}$, $\check{\mu}_\phi$, $\check{\mu}_{\phi^2}$, $\check{\sigma}^2_\phi$ , $\check{\bm{\Sigma}}_{\eta_t}$, $\check{\sigma}^2_\eta$ and $\check{\xi}_{itk}$ for $t=1,\ldots, T$; $i=1,\ldots, n_t$; $k=1,\ldots,k_i$;\\
\For{$j=1$ until convergence}{
  $\bm{\check\omega} \gets \left(\frac{\tilde{w}_{111}}{2\check\xi_{111}}\tanh(\check\xi_{111}/2),\ldots,\frac{\tilde w_{nTK}}{2\check{\xi}_{nTK}}\tanh(\check\xi_{nTK}/2)\right);$\\ 
  $\bm{\check\Omega} \gets \text{Diag}\left(\check{\bm\omega}\right);$\\ 
  $\bm{\check{\mu}}_\beta \gets (\tilde{\bm X}'\check{\bm\Omega}\tilde{\bm X} + 1/\sigma^2_\beta \bm I_p)^{-1}\tilde{\bm X}'\check{\bm\Omega}(\tilde{\bm G}\check{\bm{\mu}}_\gamma - \tilde{\bm\kappa}/\check{\bm\omega}-\tilde{\bm\Psi}\check{\bm\mu}_{\eta});$\\
  $\bm{\check{\mu}}_\gamma \gets (\tilde{\bm G}'\check{\bm\Omega}\tilde{\bm G} + 1/\sigma^2_\gamma \bm I_g)^{-1}\tilde{\bm G}'\check{\bm\Omega}(\tilde{\bm\kappa}/\check{\bm\omega}+\tilde{\bm X}\check{\bm\mu}_\beta+\tilde{\bm\Psi}\check{\bm\mu}_{\eta});$\\
  $\bm{\check{\Sigma}}_{\eta_1} \gets (\tilde{\bm\Psi}_1'\check{\bm\Omega}_1\tilde{\bm\Psi}_1 + \check{\sigma}^2_{\eta_1}(1 + \check{\mu}_{\phi^2})\bm I_m)^{-1};$\\ %
  $\bm{\check{\mu}}_{\eta_1} \gets \check{\bm \Sigma}_{\eta_1}\tilde{\bm\Psi}_1'\check{\bm\Omega}_1(\tilde{\bm G}_1\check{\bm \mu}_\gamma - \tilde{\bm\kappa}_1/\check{\bm\omega}_1 - \tilde{\bm X}_1\check{\bm\mu}_\beta) + \check{\bmu}_{\eta_2} \check{\mu}_\phi\check{\sigma}^2_{\eta_1};$ \\
  $\bm{\check{\mu}}_{\eta_1} \gets \bm{\check{\mu}}_{\eta_1} - \E[\bm{\check{\mu}}_{\eta_1}];$ //Sum-to-zero constraint\\
  \For{$t=1$ to $T$}{
    $\bm{\check{\Sigma}}_{\eta_t} \gets (\tilde{\bm\Psi}_t'\check{\bm\Omega}_t\tilde{\bm\Psi}_t + (1 + \check{\mu}_{\phi^2})\check\sigma^2_{\eta}\bm I_m)^{-1};$\\%
    $\bm{\check{\mu}}_{\eta_t} \gets \bm{\check{\Sigma}}_{\eta_t}\tilde{\bm\Psi}_t' \check{\bm\Omega}_t(\tilde{\bm G}_t\bm\check{\bmu}_\gamma - \tilde{\bm\kappa}_t/\check{\bm\omega}_t - \tilde{\bm X}_t\check{\bm\mu}_\beta) + \check{\mu}_\phi\check\sigma^2_{\eta}(\bm{\check{\mu}}_{\eta_t} + \bm{\check{\mu}}_{\eta_{t+1}}) ;$\\
  }
  $\bm{\check{\Sigma}}_{\eta_T} \gets (\tilde{\bm\Psi}_T'\check{\bm\Omega}_T\tilde{\bm\Psi}_T + \check\sigma^2_{\eta}\bm I_m)^{-1};$\\
  $\bm{\check{\mu}}_{\eta_T} \gets \bm{\check{\Sigma}}_{\eta_T}\tilde{\bm\Psi}_T' \check{\bm\Omega}_T(\tilde{\bm G}_T\bm\check{\bmu}_\gamma - \tilde{\bm\kappa}_T/\check{\bm\omega}_T - \tilde{\bm X}_T\check{\bm\mu}_\beta) + \bm{\check{\mu}}_{\eta_{T-1}}\check{\mu}_\phi\check\sigma^2_{\eta};$\\
  $\bm{\check{\mu}}_\eta \gets (\bm{\check{\mu}}_{\eta_1},\ldots,\bm{\check{\mu}}_{\eta_T}) ;$\\
  $\bm{\check{\Sigma}}_\eta \gets \text{blockdiag}(\bm{\check{\Sigma}}_{\eta_1},\ldots,\bm{\check{\Sigma}}_{\eta_T});$\\
  $\check \bSigma \gets \left(\text{blockdiag}\left(\frac{1}{\sigma^2_\gamma}I_g,\frac{1}{\sigma^2_\beta}I_q, \bm{\check{\Sigma}}_\eta\right) +\tilde{\bm C}'\bm{\check\Omega}\tilde{\bm C}\right)^{-1};$\\
    $\check\bmu \gets (\check\bmu_\gamma', \check\bmu_\beta',\check\bmu_\eta')' \gets \check\bSigma\tilde{\bm C}'\tilde{\bm\kappa};$\\
  $m \gets \sum_{t=1}^{T-1}\bm{\check{\mu}}_{\eta_{t}}'\bm{\check{\mu}}_{\eta_{t+1}}\big/\left(\sum_{t=1}^{T-1}\check{\bm\eta}_t'\check{\bm\eta}_t + \text{tr}(\check{\bSigma}_{\eta_{-T}})\right);$\\
  $\check\sigma^2_\phi\gets \check{\sigma}^2_{\eta}\big/\left(\sum_{t=1}^{T-1}\check{\bm\eta}_t'\check{\bm\eta}_t + \text{tr}(\check{\bSigma}_{\eta_{-T}})\right);$\\%
  $\ell \gets (-1 - m)/\check{\sigma}_\phi;$\\
  $u\gets (1 - m)/\check{\sigma}_\phi;$\\
  $\check\mu_\phi\gets m - \check{\sigma}_\phi\frac{\varphi(u)-\varphi(\ell)}{\Phi(u) - \Phi(\ell)} ;$\\
  $\check\mu_{\phi^2}\gets \check\mu_{\phi}^2 + \check\sigma^2_\phi\left(1-\frac{u\varphi(u)-\ell\varphi(\ell)}{\Phi(u)-\Phi(\ell)} - \left(\frac{\varphi(u)-\varphi(\ell)}{\Phi(u) - \Phi(\ell)}\right)^2\right) ;$\\
  $\check\sigma^2_{\eta_1} \gets (b + \check{\bm\mu}_{\eta_1}'\check{\bm\mu}_{\eta_1} + \text{tr}(\check{\bm\Sigma}_{\eta_1}))/(a+r/2) ;$\\ 
  $\check\sigma^2_\eta \gets (b + (1/2)(\sum_{t=2}^{T}\check{\bm\eta}_1'\check{\bm\eta}_1 +
  \text{tr}(\check{\bm\Sigma}_{\eta_{-1}} ) - 2\check{\mu}_\phi\sum_{t=1}^{T-1}\check{\bm\eta}_t\check{\bm\eta}_{t+1} +$\\
\qquad \qquad$\check{\mu}_{\phi^2}(\sum_{t=1}^{T-1}\check{\bm\eta}_t'\check{\bm\eta}_t + 
  \tr(\check{\bSigma}_{\eta_{-T}}))))/(a+r(T-1)/2);$\\
  \For{$i=1$ to $n$}{
    \For{$k=1,\ldots,k_i$}{
       \For{$t=t_i,t_{i}+1, t_{i}+2$}{
      $\check{\xi}_{ikt} \gets (\widetilde{\bm C}_{ikt}'\check\bSigma \tilde{\bm C}_{ikt} + (\tilde{\bm C}_{ikt}'\check\bmu)^2)^{1/2}$}
      }
    }
  }
\end{algorithm}
\ULforem

The final case, of a longitudinal, nominal VB model, proceeds similarly to Algorithm~\ref{alg:vb_ord_long} but without the cutpoint parameters. 
We defer this algorithm to the Appendix.

\section{Empirical Results}
\label{sec:results}
\subsection{Data description}
To illustrate the proposed methodology, we apply the ordinal models to the Household Pulse Survey (HPS).
The U.S. Census Bureau launched the HPS to measure the immediate effects of the COVID-19 pandemic on U.S. households.
The survey was first deployed during the week of April 23, 2020 and %
continues to be conducted as the Household Trends and Outlook Pulse Survey (HTOPS).\footnote{\url{https://www.census.gov/data/experimental-data-products/household-pulse-survey.html}} 
 Data collection for the survey was split into ``phases'' of varying lengths, which differed in methodology, data collection, and structure.
The intention was to make it straightforward to modify the survey and add new questions as needed amid the uncertainty of the pandemic.

Though many questions ran for the duration of HPS, the first phase, which lasted 12 weeks, was unique in its longitudinal structure.
During this time, households were repeatedly interviewed according to a rotating panel design.
Respondents were selected in an initial week, then re-interviewed for up to two more weeks if they continued to provide responses but dropped from the sampling frame if they failed to provide a response at any point.
Therefore, each respondent provided between one and three consecutive responses.
We focus on applying our methods to this first phase because of its longitudinal nature.
It is important to note, however, that any modeling of later HPS phases that includes the first 12 weeks will still require using our methodology for these early measurements.
Modeling of later phases also still requires capturing the temporal dependence due to the repeated cross-sectional design as our model does.
HTOPS is intended to reinstate the longitudinal design, as well.

Collaboration with numerous federal agencies led to a range of survey items in the HPS.
These span topics like economic well-being, physical and mental health, and other demographic information.
A large number of these questions are categorical in nature.
For instance, %
the Bureau of Labor Statistics included questions about the receipt and use of the Economic Impact Payment stimulus (e.g., was it used to (1) pay for expenses, (2) pay off debt, (3) add to savings, or (4) not applicable).\footnote{\url{bls.gov/cex/research_papers/pdf/safir-effects-of-covid-on-household-finances.pdf}}
The Health Resources and Services Administration asked questions on childcare and children's health (e.g., ``whether any children in the last 4 weeks showed any of [\ldots] 8 mental health-related behaviors'').\footnote{\url{https://mchb.hrsa.gov/covid-19/data}}
The National Center for Health Statistics (NCHS) added questions regarding delayed medical care 
 and anxiety and depression symptoms (e.g., frequency (1) not at all, (2) several days, (3) more than half the days, or (4) nearly every day).\footnote{\url{https://www.cdc.gov/nchs/covid19/health-care-access-and-mental-health.htm}} %
 As such, producing precise estimates for HPS questions is of use to a large number of stakeholders and requires modeling longitudinal categorical data.

In particular, questions regarding mental health are of interest to researchers and policymakers due to the unprecedented disruptions to daily life, working conditions, and mental health services associated with the pandemic and the concomitant public health measures \citep{WHO_2022}. 
Longitudinal studies are important here in order to judge the effect on individuals over time and to determine if changes are ``acute'' or sustained \citep{daly_2022}.
Also, it is posited that specific demographic groups, such as women and young adults, are disproportionately impacted \citep{hawes_2022, metin_2022}.
This makes our method suited to the task. 
Design-based estimates, on the other hand, break down for group comparisons in finer partitions since the standard error estimates will be unstable for subdomains with low sample sizes and many domains will have no sampled units at all.
For this reason, we focus our data analysis in Section~\ref{subsec:data_analysis} on the aforementioned NCHS question regarding anxiety.

\subsection{Simulation study}
To assess the performance of our ordinal models, we conduct an empirical simulation study,
for the aforementioned response regarding frequency of anxiety over course of the week, which is ranked on an ascending scale of 4 categories. 
We take as our ground-truth population all $774,882$ unique respondents and $991,412$ total responses in the HPS sample from the continental United States (excluding the District of Columbia).

To mimic a true complex survey, we draw informative subsamples from this population and obtain model-based and direct estimates for each subsample.
We compare based on summary measures, including mean square error (MSE), absolute bias, credible interval coverage rate, and the interval score (IS) of \citet{Gneiting_2007}.
The interval score is calculated as
\[IS_{\alpha}(\ell, u; x) = (u-\ell)+\frac{2}{\alpha}(\ell -x){1}\{x< \ell\} + \frac{2}{\alpha}(x-u){1}\{x>u\}\]
and penalizes overly wide predictive intervals.
A lower score is more desirable.

To make the subsamples informative, we take a probability proportional to size sample following the Poisson method of \citet{brewer_early_hanif_1984} with an expected sample size of $5\%$ of the population.
Each member of the population is sampled with a probability $\pi_i$ proportional to a size variable
\[\pi_i \propto \exp\{.1w_i^* + .2\bar{y}_i \},\]
where $w_i^*$ is the log survey weight of household $i$, scaled to have zero mean and standard deviation one, and $\bar{y}_i$ the scaled mean of all of a household's responses.
The weight assigned to sampled unit $i$ is then $w_i=1/\pi_i.$

Spatial basis functions are constructed via the process outlined in Section~\ref{subsec:spatio-temporal}. 
Retaining only the eigenvectors corresponding to positive eigenvalues leads to basis functions of dimension $m=20$.
Model covariates include race (white, Black, Asian, or other), sex (male, female), and age categories (one category for 18-25, one for 65+, and the remaining intermediary ages binned into five-year groups, for a total of 10 categories). 
The design matrix includes no intercept and further drops a reference category from one of the covariates in order for the cutpoints to be identifiable.
To capture longitudinal dependence in the nominal model, we include a ``synthetic'' covariate that indexes a household's previous response category (or lack thereof).
In the ordinal model, this variable is used to vary the cutpoints as described in Section~\ref{subsec:spatio-temporal}.%

We compute direct estimators, and fit cross-sectional and longitudinal variants of both the Gibbs and VB models for a total of five different estimators.
 The cross-sectional models are fit independently for each week.
For each of the models, $R$ posterior draws (or draws from the variational distribution)
 of the parameter estimates are taken.
For all models, we set $R=1,500$, and discard $500$ burn-in iterations beforehand in the Gibbs sampler. 
The estimated probability for a given category $k=1,\ldots, K-1$ is calculated as
\[\tilde\pi_{itk}^{(r)} = \logit^{-1}(\bm{u}'_{itk}\bm{\gamma}^{(r)} - \bm x_i'\bm\beta_k^{(r)} - \bm\psi_i\bm\eta_t^{(r)}).\]
The full probability vector $\hat{\bm\pi}_{it}=(\hat{\pi}_{it1},  \ldots,\hat{\pi}_{itK})$ is reconstructed from the stick-breaking representation and estimates for the full population are then generated as
\[\hat{y}_{it}^{(r)}\mid\cdot \sim \text{Multinomial}(\bm n, \bm \pi_i)\]
for $t=1,\ldots, T$ and $i = 1,\ldots, N$, where $\bm n$ is a K-dimensional vector of ones.
This population can then be aggregated into cells by taking the average 
\[\hat{\bar y}_{jtk}^{(r)} = \frac{1}{N_j}\sum_{i\in \mathcal{S}_j} \hat{y}_{itk}^{(r)}\]
for each subdomain $j$
and summary metrics are computed over the posterior distributions of the cell estimates.

The results of $100$ simulation runs are summarized in Table~\ref{tab:ord_sim}.
Regardless of sampling algorithm, the longitudinal models lead to improved MSE, lower interval scores, and faster runtimes as compared to their cross-sectional counterparts.
The Gibbs models also outperform their VB counterparts on MSE and IS, but with the trade-off of a greatly increased runtime.
The longitudinal Gibbs model performs best overall in terms of MSE and interval score.
Although the direct estimator attains nearly the nominal coverage rate, it does so at the cost of overly wide confidence intervals, hence its high interval score.
\begin{table}
\centering
\caption{Overall results for direct- and model-based estimates averaged over 100 simulations in the ordinal case. Runtime is the median of 100 runs. All other values are averages over weeks, areas, and simulations.\label{tab:ord_sim}}%
\begin{tabular}{lrrrrr}
  \hline
Method & MSE & Abs Bias & Cov. & IS & Runtime (s)\\
  \hline
   Direct & $2.7 \times 10^{-3}$ & ${4.0 \times 10^{-3}}$ & ${95\%}$ & $.252$ & - \\
    VB-CS & $5.4 \times 10^{-4}$ & $1.4 \times 10^{-2}$ & $89\%$ & $.115$ & 15\\
    
   VB-Lon & $4.1 \times 10^{-4}$ & $1.4 \times 10^{-2}$ & $86\%$ & $.113$ & $43$\\
 Gibbs-CS & $5.1 \times 10^{-4}$ & $1.4 \times 10^{-2}$ & $90\%$ & $.112$ & $99,103$\\
Gibbs-Lon & $\mathbf{3.7 \times 10^{-4}}$ & $1.3 \times 10^{-2}$ & $89\%$ & $\mathbf{.099}$ & $15,673$\\
  \hline
\end{tabular}
\end{table}

\begin{figure}[ht!]
    \centering
    \includegraphics[width=\linewidth]{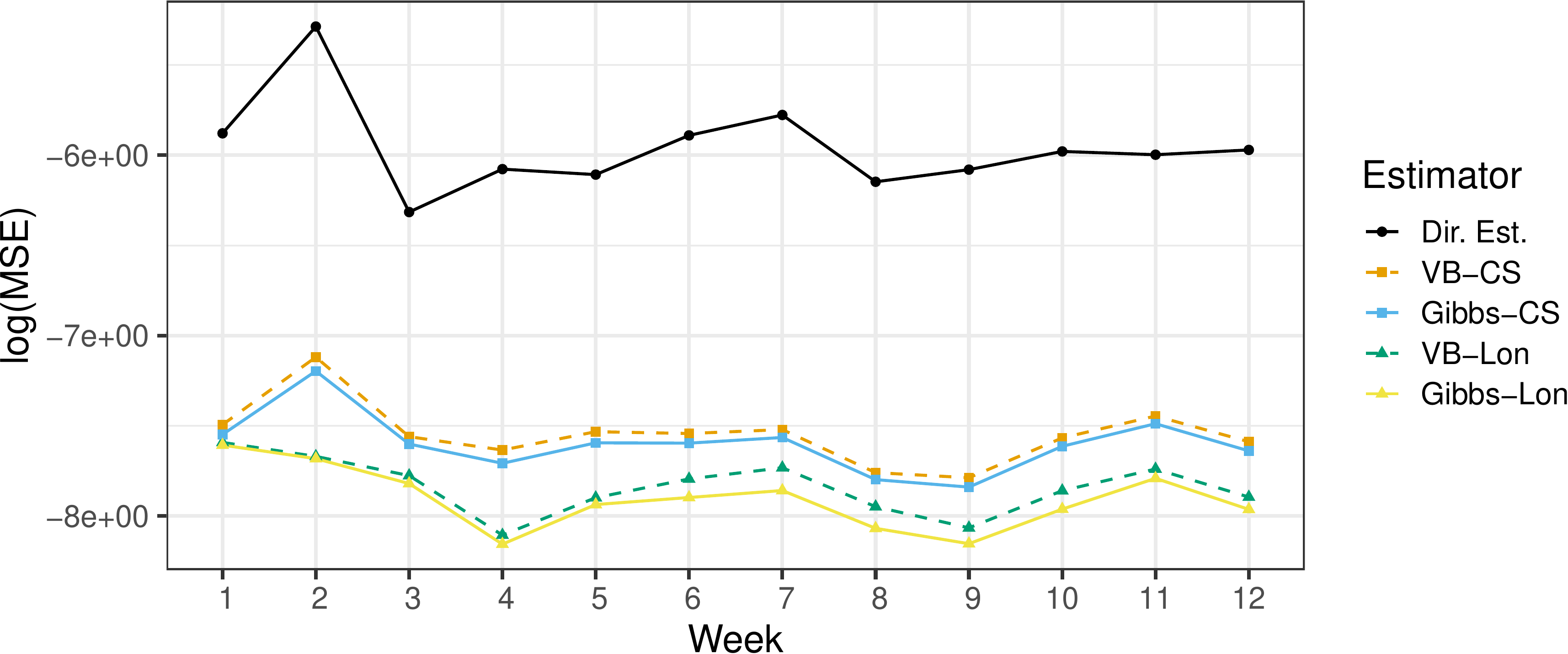}
    \caption{Time series plot of the MSE for each estimator by week in the ordinal empirical simulation study.
             MSE is calculated for every combination of week and area then averaged across areas and 100 samples.}
    \label{fig:emp_sim_ordinal_MSE_time}
\end{figure}

For a more detailed look, Figure~\ref{fig:emp_sim_ordinal_MSE_time} shows a time series plot of the MSE for each estimator averaged over areas and simulations.
Again, all models outperform the direct estimator, and we see the same pattern.
Longitudinal models uniformly reduce the MSE over their cross-sectional counterparts and the Gibbs models uniformly reduce MSE over their VB counterparts, with the longitudinal Gibbs sampler displaying the lowest MSE.

\begin{figure}[ht!]
    \centering
    \includegraphics[width=\linewidth]{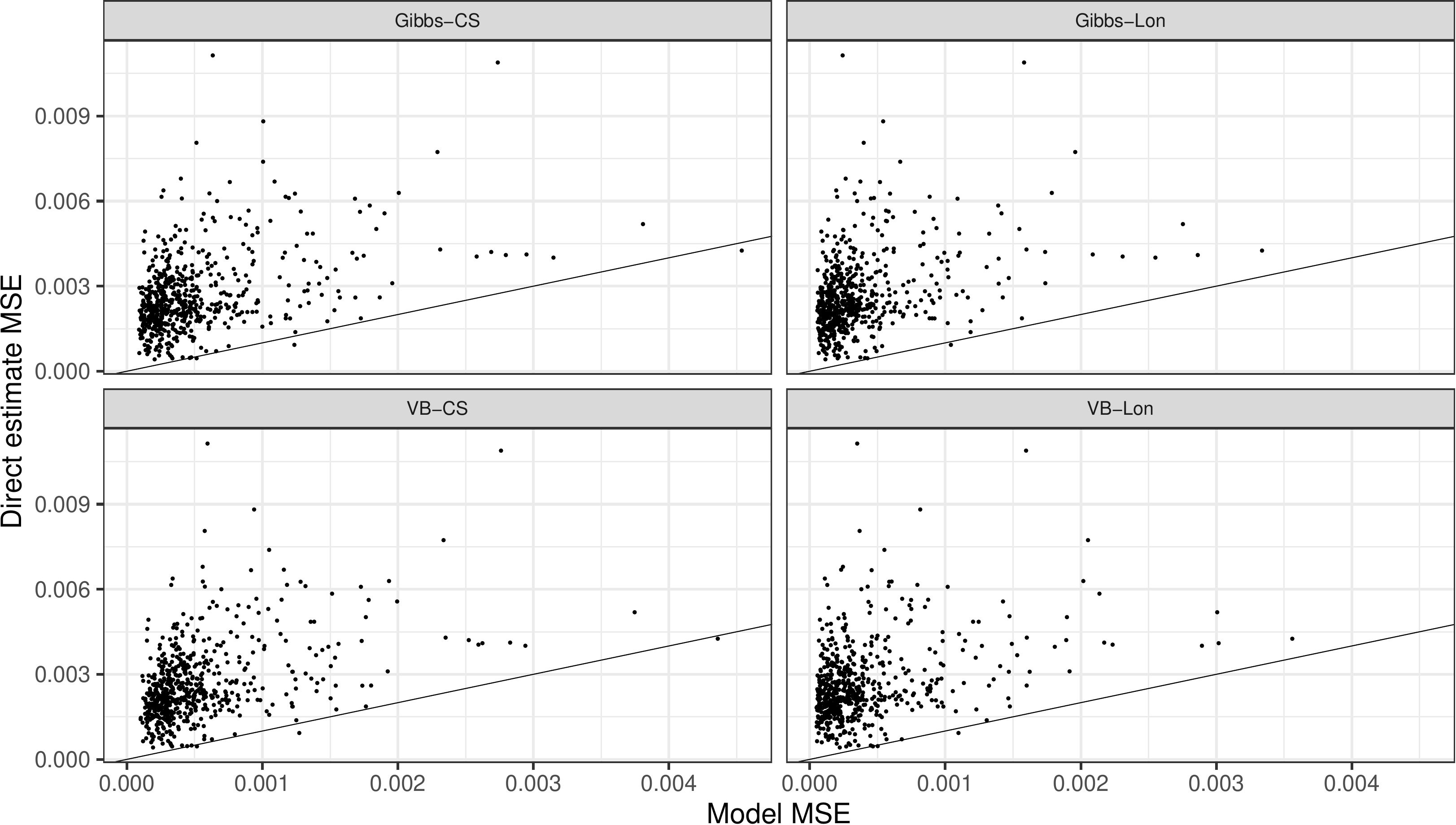}
    \caption{Ratio of model MSE to direct estimate MSE for each of the ordinal models in each area-week post-stratification cell averaged across 100 simulations.
      Values less than one indicate a reduction in MSE over the direct estimate.}
    \label{fig:ordinal_MSE_ratios}
\end{figure}

As a further diagnostic, Figure~\ref{fig:ordinal_MSE_ratios} plots the ratios of the model MSE for a given time-area combination relative to the direct estimator.
The $x=y$ reference line denotes a ratio of one, indicating identical MSE between a model and the direct estimator.
Only a few domains have a ratio less than one, with the longitudinal Gibbs model having the fewest.

\subsection{Analysis of Household Pulse Anxiety Data}
\label{subsec:data_analysis} 
We conduct a data analysis by running the ordinal, longitudinal VB algorithm on the entire HPS dataset for the same anxiety response considered in the simulation study. 
For population values of the design matrix we use data from the 2020 Census Population Estimates program.\footnote{Available at \url{https://www2.census.gov/programs-surveys/popest/datasets/2020-2021/state/asrh/sc-est2021-alldata6.csv}}
The choice of covariates and basis functions is the same as in the simulation studies.

\begin{figure}[h]
    \centering
    \includegraphics[width=\linewidth]{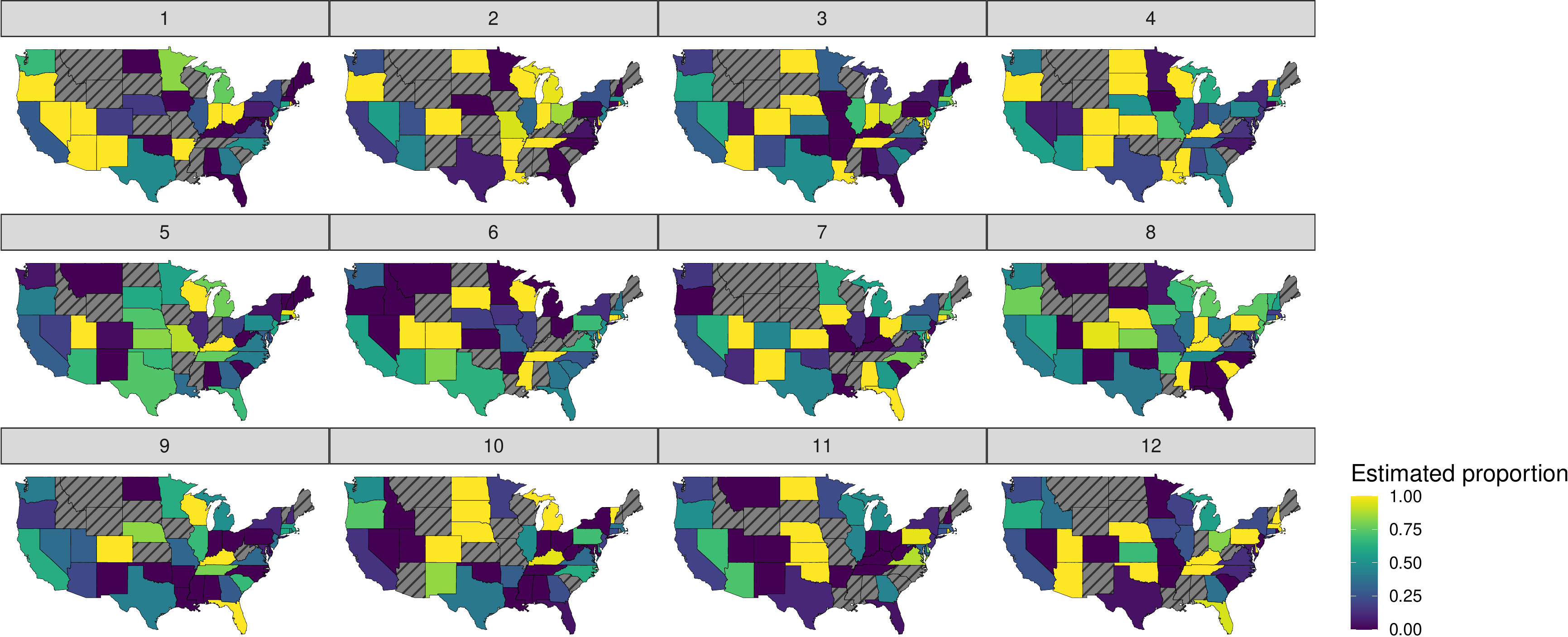}
    \caption{Areal time series plot of the direct estimates for proportion of Asian males aged 36 to 40, who feel ``not at all" anxious during the past week. Estimates are calculated from the entire HPS Phase 1 data.  The plot is faceted by week, with the  first time point at top-left and the last time-point at bottom-right.}
    \label{fig:ordinal_dir_est_time_series}
\end{figure}

To emphasize our models' ability to make predictions for small areas, we consider a finer partition into subdomains and present results for a particular combination of race, sex, and age -- namely, Asian males aged 36 to 40 -- rather than weekly area-level averages as in the simulation.
This requires first making estimates for a cross-tabulation of two sex categories, four race categories, 10 age categories and the $K$ response categories across 49 states and 12 weeks.
For the given ordinal response with $K=4$ there are $94,080$ domains to estimate.
Relying on direct estimators is of little use in this context, where $41\%$ of the domains have zero respondents, only $3\%$ of cells have a sample size greater than 30, and the median sample size for all cells is one.
The effect of this sparsity is reflected in Figure~\ref{fig:ordinal_dir_est_time_series} which plots the direct estimates for a single response category among Asian males between the ages of 36 and 40.
Many states have no estimate and different states are missing estimates in different weeks, meaning uninterrupted longitudinal estimates are generally unobtainable.
Where the direct estimator can be calculated, the resulting values tend to be at the boundary, either zero or one.

\begin{figure}[h]
    \centering
    \includegraphics[width=\linewidth]{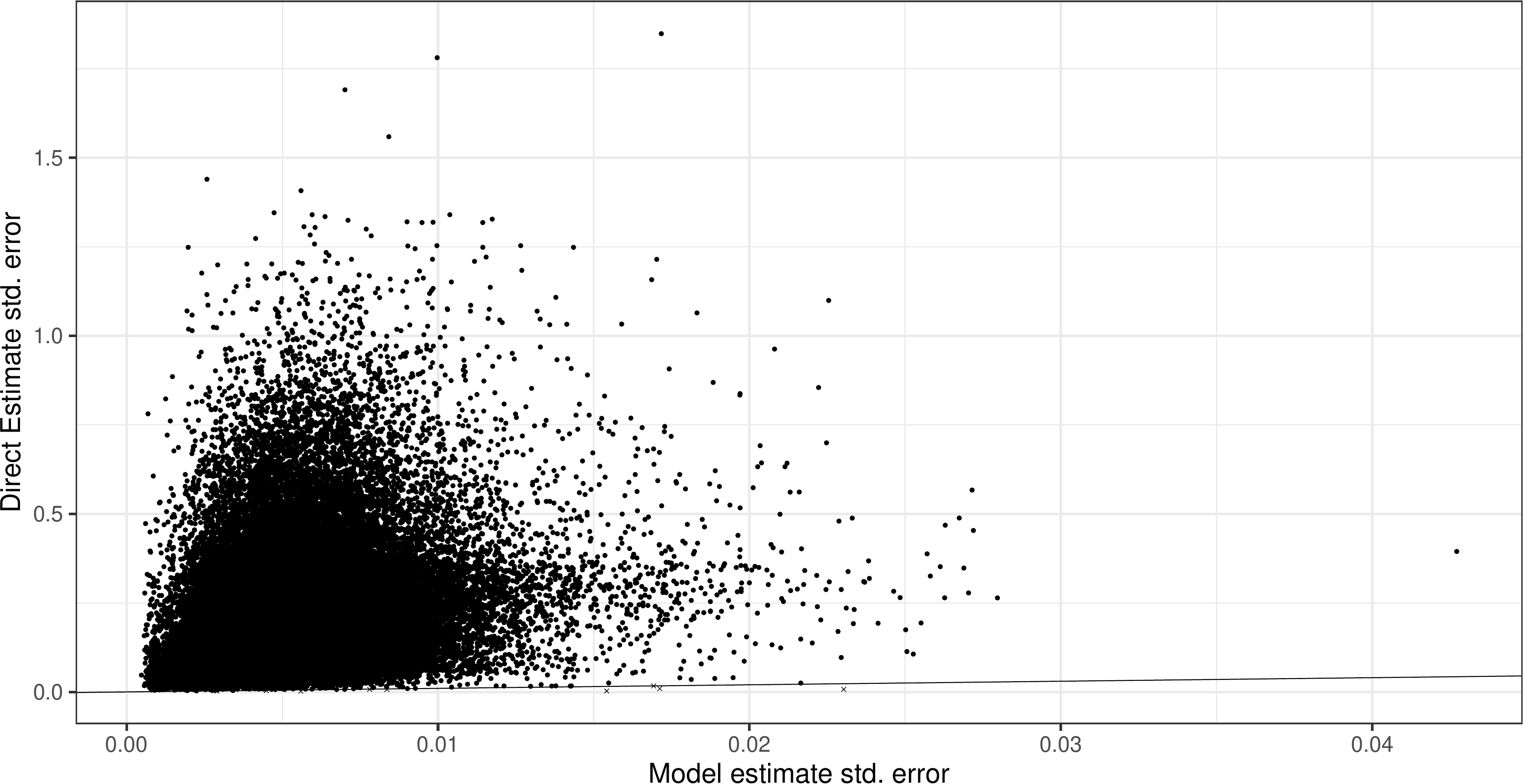}
    \caption{Scatter plot of ratio of model standard errors to direct estimate standard errors. Values less than one indicate an improvement of the model over the direct estimator. The four values less than one are marked with crosses.}
    \label{fig:ordinal_std_err_ratio}
\end{figure}

The standard errors of the direct estimates are calculated using the replicate weights method in the R \textit{survey} library \citep{R_survey}.
Like the point estimates, these values tend to be extreme and the model estimates improve on them substantially.
In fact, for $60\%$ of domains, a valid standard error estimate cannot be produced for the direct estimator, either because the sample size is too low or because all response values in a given cell are identical.
For the remaining cells where direct estimate standard errors can be produced, plotting the ratio of model standard errors to those of the direct estimate in Figure~\ref{fig:ordinal_std_err_ratio} demonstrates a reduction in all but nine domains ($.01\%$ of the total).

\begin{figure}[h]
    \centering
    \includegraphics[width=\linewidth]{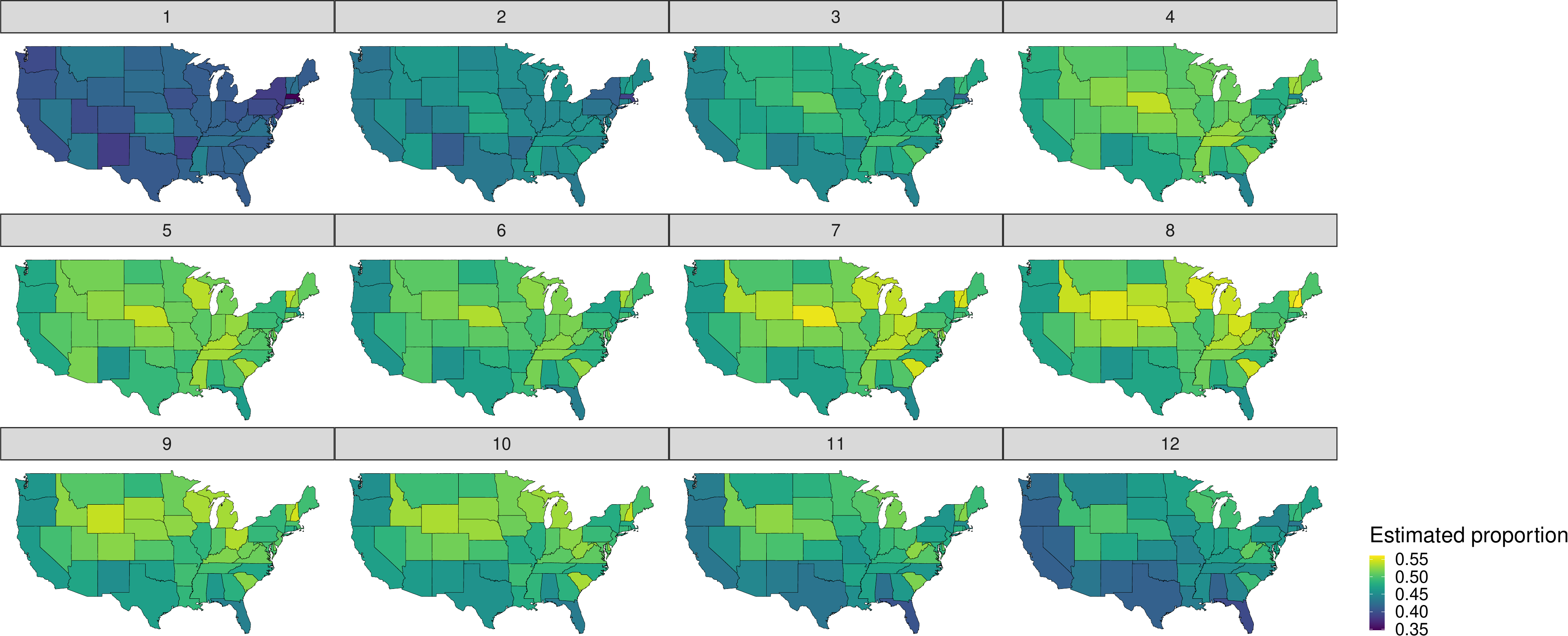}
    \caption{Areal time series plot of the ordinal VB model estimates for proportion of Asian males aged 36 to 40, who feel ``not at all" anxious during the past week. Estimates are calculated from the entire HPS Phase 1 data.  The plot is faceted by week, with the  first time point at top-left and the last time-point at bottom-right.}
    \label{fig:ordinal_model_time_series}
\end{figure}

Moreover, model estimates in Figure~\ref{fig:ordinal_model_time_series} show much smoother trends across both space and time whereas the direct estimates in Figure~\ref{fig:ordinal_dir_est_time_series} jump around dramatically.
For example, the estimated proportion for Florida is nearly zero in week 8, nearly one in the following week, then back to zero in week 10.
The underlying truth is highly unlikely to change so rapidly.
A similar pattern can be observed for a number of other states, such as Iowa, which jumps between two extreme values in weeks six and seven, with the additional complication that many weeks also have no estimate so that the trend is unobservable.

\begin{sidewaysfigure}
    \centering
    \includegraphics[width=\linewidth]{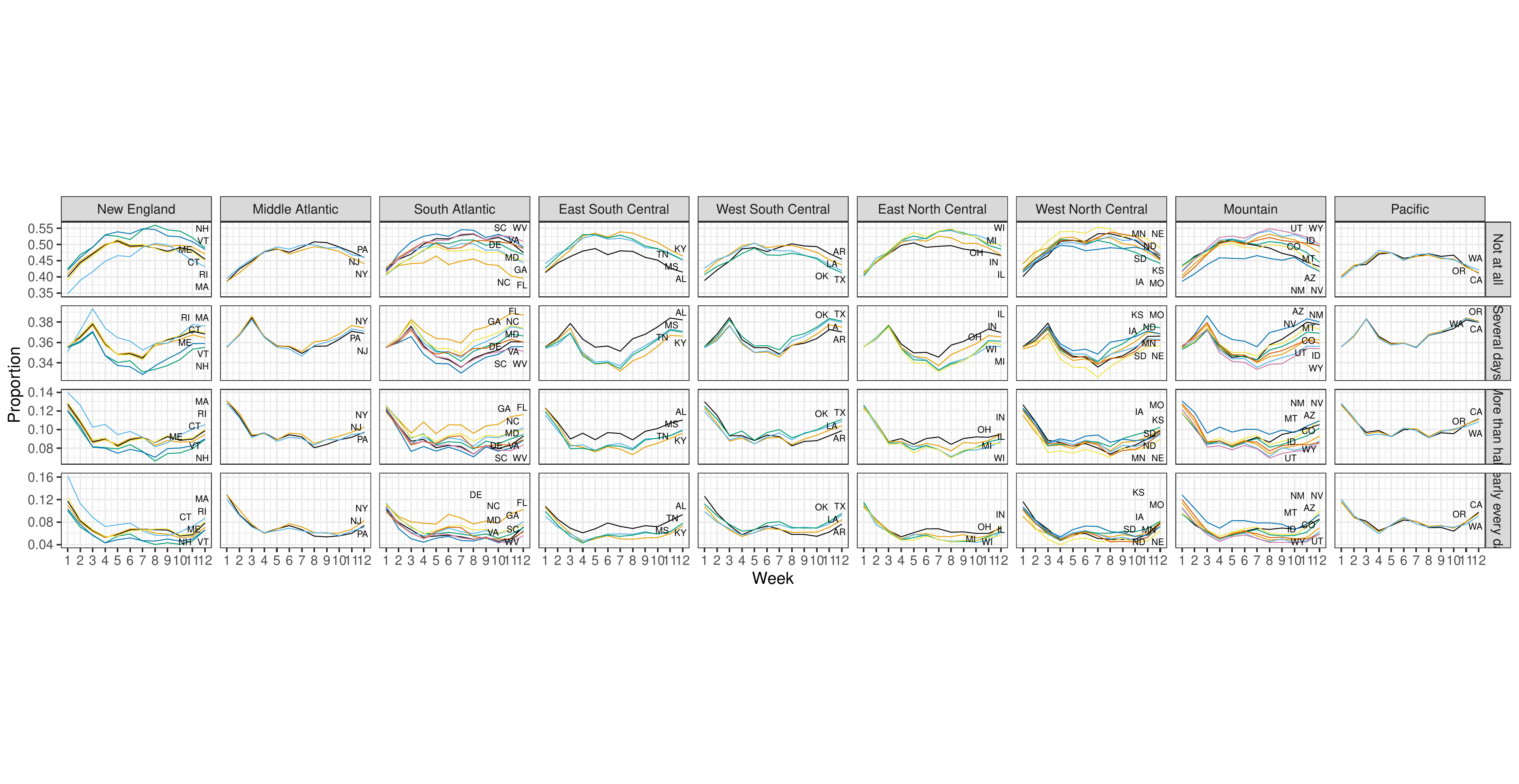}
    \caption{Longitudinal VB model estimates for frequency of anxiety over the past 7 days for Asian males aged 36 to 40 grouped by Census division.}
    \label{fig:ordinal_divisions}
\end{sidewaysfigure}

Additionally, if we want to compare multiple response categories simultaneously, our unit-level modeling approach allows us to visualize detailed estimates as in Figure \ref{fig:ordinal_divisions}, which plots estimated trajectories for the proportion of each category in a given demographic group by state. 
For readability, the states are grouped into the nine geographical Census divisions.
The first timepoint of the plots corresponds to April 23, 2020 and the last corresponds to July 21. %
Over the course of this time period the proportion of respondents reporting no anxiety at all grows and then decreases.
In week two, all states exhibit an increase in the proportion of ``several days'' of anxiety while ``nearly every day'' stays flat until an uptick occurs, starting in week eight.
We can also see that some Census divisions, like the Pacific, are fairly homogeneous.
In contrast, the Mountain division shows more variability and Nevada, whose economy is far more dependent on tourism, is estimated to have distinctly more anxiety than its neighbors.

\begin{figure}
    \centering
    \includegraphics[width=\linewidth]{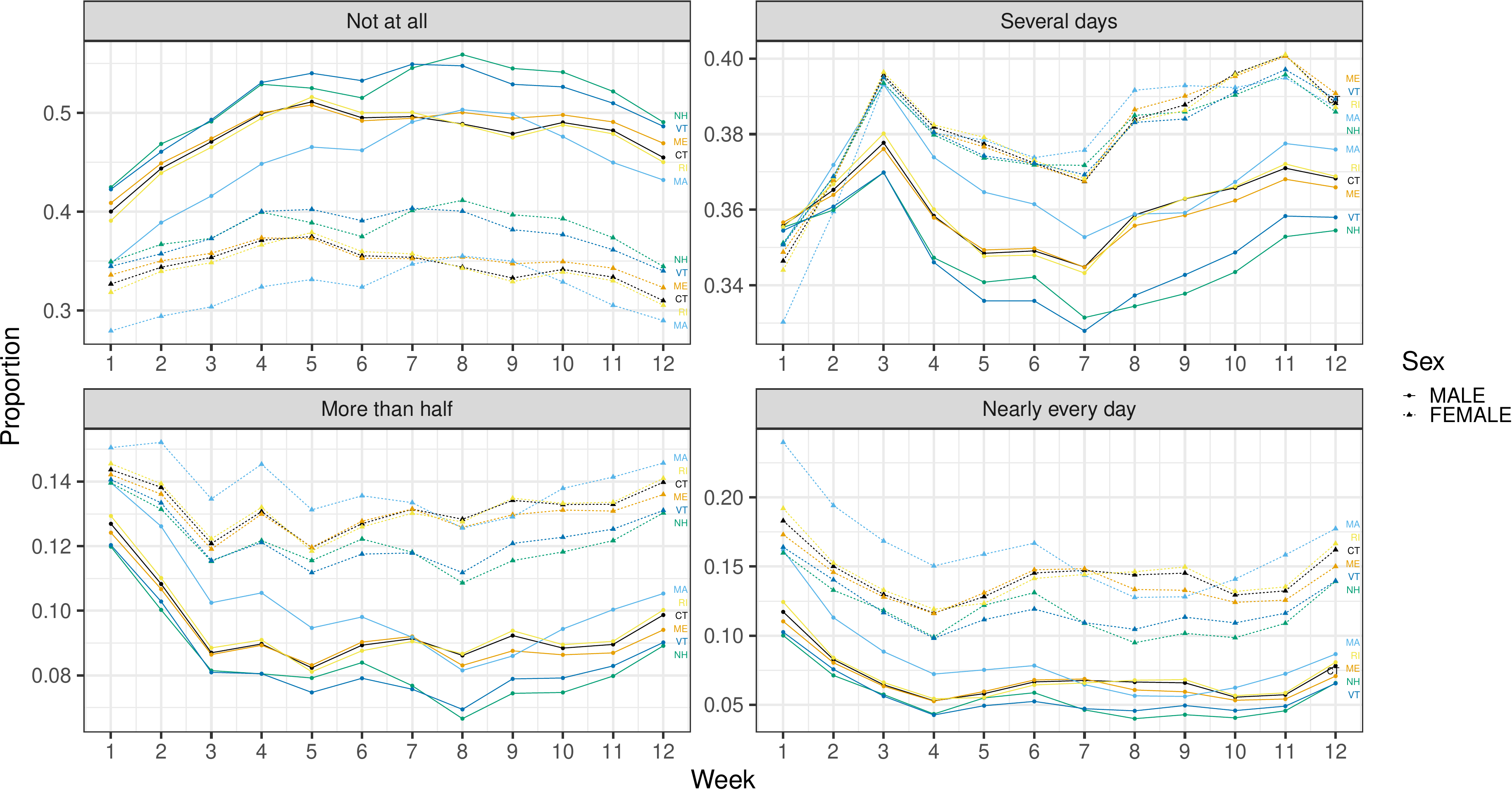}
    \caption{Longitudinal VB model estimates for frequency of anxiety over the past 7 days comparing Asian males and females aged 36 to 40 in the New England Census division.}
    \label{fig:ordinal_sex_comparison}
\end{figure}

Of further interest are group comparisons, which the direct estimator would be incapable of handling, since it considers each group independently. 
Figure~\ref{fig:ordinal_sex_comparison} shows a comparison between the trajectories for men and women in a single Census division.
The trends between the two groups are similar across states, but women are more likely to report some frequency of anxiety and less likely to report feeling not at all anxious.
This is in line with existing research \citep{metin_2022}.
Similar patterns hold for the other
Census divisions. %

\section{Discussion}
\label{sec:discuss}

In this paper, we have introduced Bayesian models for both nominal and ordinal survey data and both cross-sectional and longitudinal designs, where a special case of the cross-sectional ordinal model is a conjugate Bayesian, ordinal logistic regression.
Alongside the work of \citet{parker_holan_janicki_2022} and \citet{vedensky_2023}, this covers all of the most common types of survey responses within a unified, unit-level modeling framework.
The models are efficient, capture dependencies that area-level models are not able to, and lead to great improvements over direct estimators  as shown in the simulation study. 
The application to HPS phase 1 data regarding mental health symptoms demonstrates the types of fine-grained comparisons we are able to make that would not be possible in an area-level approach.

The longitudinal models in this paper extend readily to a number of large and important surveys that follow rotating panel designs such as the Survey of Income and Program Participation or the National Crime Victimization Survey. 
Future work remains to mitigate some of the biases inherent to these designs such as rotation group bias \citep{bailar1975} and to
consider other longitudinal designs, possibly with uneven sampling intervals.%
\section*{Acknowledgements}
This article is released to inform interested parties of ongoing research and to encourage discussion. The views expressed on statistical issues are those of the authors and not those of the NSF or U.S. Census Bureau.

\section*{Funding}
This research was partially supported by the Census Bureau Dissertation Fellowship program and the U.S. National Science Foundation (NSF) under NSF grants NCSE-2215168, and NCSE-2215169.

\section*{Appendix}
MCMC algorithms for the cross-sectional ordinal and longitudinal ordinal are presented below, as is the VB algorithm for longitudinal, nominal data.. 
For details on the cross-sectional nominal MCMC and VB algorithms, see \citet{parker_holan_janicki_2022}. 
\subsection*{Gibbs sampling algorithms} 
\subsubsection*{Cross-sectional ordinal}
\begin{enumerate}
    \item Sample $\omega_{ik} | \cdot \sim \text{PG}(\tilde{w}_{i}n_{ik}, \bm u_{ik}' - \bm x_i'\bm\beta -\bm\psi_i'\bm\eta)$ for $i=1,\ldots, n$ and $k=1,\ldots,k_i$
    \item Sample $\sigma^2_{\eta}|\cdot \sim \text{IG}(a+m/2, b+\bm\eta'\bm\eta/2)$
    \item Sample $\bm\beta|\cdot \sim \text{N}_p\left(\bm\mu=\left(\bm X'\bm\Omega(G'\bm\gamma-\bm\kappa/\bm\omega-\bm\Psi\bm\eta)\right), \bm\Sigma=\left(\bm X'\bm\Omega\bm X + \frac{1}{\sigma^2_\beta}I_p\right)^{-1}\right)$
    \item Sample $\bm\gamma |\cdot \sim \text{N}_g\left(\bm\mu=\left(\bm G'\bm\Omega(X'\bm\beta-\bm\kappa/\bm\omega-\bm\Psi\bm\eta)\right), \bm\Sigma=\left(\bm G'\bm\Omega\bm G + \frac{1}{\sigma^2_\gamma}I_g\right)^{-1}\right)$
    \item Sample $\bm\eta|\cdot \sim \text{N}_m\left(\bm\mu=\left(\bm {\Psi}'\bm\Omega(X'\bm\beta-\bm\kappa/\bm\omega-\bm\Psi\bm\eta)\right), \bm\Sigma=\left(\bm{\Psi}'\bm\Omega\bm \Psi + \frac{1}{\sigma^2_{\eta_1}}I_g\right)^{-1}\right)$
\end{enumerate}

\subsubsection*{Longitudinal ordinal}
Below, $\Psi_f$ is the matrix of basis functions for all areas and timepoints, corresponding to $\text{vec}(\bm\eta)$.
\begin{enumerate}
    \item Sample $\omega_{itk}|\cdot$ for $i=1,\ldots, n_t$, $t=1, \ldots, n_t$, $k=1,\ldots,k_i$
       \[\omega_{itk} | \cdot \sim \text{PG}(\tilde{w}_{it}n_{itk}, \bm u_{itk}' - \bm x_i'\bm\beta -\bm\psi_i'\bm\eta_t) \]
    \item Sample $\phi |\cdot \sim \text{TruncNorm}\left(\frac{\sum_{t=2}^T\bm\eta_t'\bm\eta_{t-1}}{\sum_{t=2}^T\bm\eta_{t-1}'\bm\eta_{t-1}},\frac{\sigma^2_\eta}{\sum_{t=2}^T\bm\eta_{t-1}'\bm\eta_{t-1}},-1,1\right)$
    \item Sample $\sigma^2_{\eta_1}|\cdot \sim \text{IG}(a+m/2, b+\frac{1}{2}\bm\eta_1'\bm\eta_1)$
    \item Sample $\sigma^2_{\eta}|\cdot \sim \text{IG}(a+ m(T-1)/2, b+\sum_{t=2}^T(\bm\eta_t-\phi\bm\eta_{t-1})'(\bm\eta_t-\phi\bm\eta_{t-1})/2)$
    \item Sample $\bm\beta:$
    \[\bm\beta|\cdot \sim \text{N}_p\left(\bm\mu=\left(\bm \tilde{\bm X}'\bm\Omega(\tilde{\bm{G}}\bm\gamma-\tilde{\bm\kappa}/\bm\omega-\tilde{\bm\Psi}_f\text{vec}(\bm\eta))\right), \bm\Sigma=\left(\tilde{\bm X}'\bm\Omega\tilde{\bm X} + \frac{1}{\sigma^2_\beta}I_p\right)^{-1}\right)\]
    \item Sample $\bm\gamma:$
    \[\bm\gamma|\cdot \sim \text{N}_g\left(\bm\mu=\left(\tilde{\bm G}'\bm\Omega(\tilde{\bm\kappa}/\bm\omega+\tilde{\bm X}'\bm\beta+\tilde{\bm\Psi}_f\text{vec}(\bm\eta)\right), \bm\Sigma=\left(\tilde{\bm G}'\bm\Omega\tilde{\bm G} + \frac{1}{\sigma^2_\gamma}I_g\right)^{-1}\right)\]
    \item Sample $\bm\eta_1:$
    \[\bm\eta_1|\cdot \sim \text{N}_m\left(\bm\mu=\left(\tilde{\bm \Psi}_1'\bm\Omega_1(\tilde{\bm G}_1'\bm\gamma - \tilde{\bm\kappa}_1/\bm\omega_1 - \tilde{\bm X}_1\bm\beta)\right), 
    \bm\Sigma=\left(\tilde{\bm \Psi}_1'\bm\Omega_1\tilde{\bm \Psi}_1 + \left(\frac{1}{\sigma^2_{\eta_1}}+\frac{\phi^2}{\sigma^2_{\eta}}\right)I_m\right)^{-1}\right)\]
    \item Sample $\bm\eta_t$, for $t=2,\ldots, T-1:$
    \[\bm\eta_t |\cdot \sim \begin{aligned}[t]\text{N}_p\Bigg(\bm\mu &= \bm\Sigma \left(\tilde{\bm{ \Psi}}_t'\bm\Omega_t(\tilde{\bm G}'_t\bm\gamma-\tilde{\bm\kappa}_t/\bm\omega_t - \tilde{\bm{X}}_t\bm\beta)+\frac{\phi}{\sigma^2_\eta}(\bm\eta_{t-1}+\bm\eta_{t+1})\right),\\                                   
                                         \bm\Sigma&=\left(\tilde{\bm{\Psi}}_t'\bm\Omega_t\tilde{\bm{\Psi}}_t + \frac{1+\phi^2}{\sigma^2_\eta}I_m\right)^{-1}\Bigg)\end{aligned}\]
    \item Sample $\bm\eta_T$
    \[\bm\eta_T |\cdot \sim \text{N}_p\left(\bm\mu = \bm\Sigma \left( \tilde{\bm{\Psi}}_T'\bm\Omega_T(\bm {\tilde \bm G}_T'\bm\gamma_T - \tilde{\bm\kappa}_T/\bm\omega_T - \tilde{\bm{X}}_T\bm\beta)+\frac{\phi}{\sigma^2_\eta}\bm\eta_{T-1}\right),                                                                        
                                          \bm\Sigma=\left(\tilde{\bm{\Psi}}_T'\bm\Omega_T\tilde{\bm{\Psi}}_t + \frac{1}{\sigma^2_\eta}I_m\right)^{-1}\right)\]
\end{enumerate}

\subsection*{Longitudinal, nominal VB algorithm}
For longitudinal, nominal data, the binary VB algorithm below can be fit $K-1$ times to data that is factored according to the stick-breaking representation (\ref{eq:mult_sb}).
\normalem
\begin{algorithm}[ht!]
\small
\caption{Variational Bayes algorithm for the longitudinal binary model\label{alg:vb_nom_long}}
\KwData{${\bm C}=[{\bm X}, {\bm \Psi}]$ and ${\bm\kappa}$}
Initialize $\check{\bm{\mu}}_{\eta}$, $\check{\mu}_\phi$, $\check{\mu}_{\phi^2}$, $\check{\sigma}^2_\phi$ , $\check{\bm{\Sigma}}_{\eta_t}$, $\check{\sigma}^2_\eta$ and $\check{\xi}_{itk}$ for $t=1,\ldots, T$; $i=1,\ldots, n_t$;\\
\For{$j=1$ until convergence}{
  $\bm{\check\omega} \gets \left(\frac{\tilde{w}_{11}}{2\check\xi_{11}}\tanh(\check\xi_{11}/2),\ldots,\frac{ \tilde w_{n_TT}}{2\check{\xi}_{n_TT}}\tanh(\check\xi_{n_TT}/2)\right);$\\ 
  $\bm{\check\Omega} \gets \text{Diag}\left(\check{\bm\omega}\right);$\\ 
  $\bm{\check{\mu}}_\beta \gets ({\bm X}'\check{\bm\Omega}{\bm X} + 1/\sigma^2_\beta \bm I_p)^{-1}{\bm X}'\check{\bm\Omega}({\bm\kappa}/\check{\bm\omega}-{\bm\Psi}\check{\bm\mu}_{\eta});$\\
  $\bm{\check{\Sigma}}_{\eta_1} \gets ({\bm\Psi}_1'\check{\bm\Omega}_1{\bm\Psi}_1 + \check{\sigma}^2_{\eta_1}(1 + \check{\mu}_{\phi^2})\bm I_m)^{-1};$\\ %
  $\bm{\check{\mu}}_{\eta_1} \gets \check{\bm \Sigma}_{\eta_1}{\bm\Psi}_1'\check{\bm\Omega}_1({\bm\kappa}_1/\check{\bm\omega}_1 - {\bm X}_1\check{\bm\mu}_\beta) + \check{\bmu}_{\eta_2} \check{\mu}_\phi\check{\sigma}^2_{\eta_1};$ \\
  $\bm{\check{\mu}}_{\eta_1} \gets \bm{\check{\mu}}_{\eta_1} - \E[\bm{\check{\mu}}_{\eta_1}];$\\
  \For{$t=1$ to $T$}{
    $\bm{\check{\Sigma}}_{\eta_t} \gets ({\bm\Psi}_t'\check{\bm\Omega}_t{\bm\Psi}_t + (1 + \check{\mu}_{\phi^2})\check\sigma^2_{\eta}\bm I_m)^{-1};$\\%
    $\bm{\check{\mu}}_{\eta_t} \gets \bm{\check{\Sigma}}_{\eta_t}{\bm\Psi}_t' \check{\bm\Omega}_t({\bm\kappa}_t/\check{\bm\omega}_t - {\bm X}_t\check{\bm\mu}_\beta) + \check{\mu}_\phi\check\sigma^2_{\eta}(\bm{\check{\mu}}_{\eta_t} + \bm{\check{\mu}}_{\eta_{t+1}}) ;$\\
  }
  $\bm{\check{\Sigma}}_{\eta_T} \gets ({\bm\Psi}_T'\check{\bm\Omega}_T{\bm\Psi}_T + \check\sigma^2_{\eta}\bm I_m)^{-1};$\\
  $\bm{\check{\mu}}_{\eta_T} \gets \bm{\check{\Sigma}}_{\eta_T}{\bm\Psi}_T' \check{\bm\Omega}_T({\bm\kappa}_T/\check{\bm\omega}_T - {\bm X}_T\check{\bm\mu}_\beta) + \bm{\check{\mu}}_{\eta_{T-1}}\check{\mu}_\phi\check\sigma^2_{\eta};$\\
  $\bm{\check{\mu}}_\eta \gets (\bm{\check{\mu}}_{\eta_1},\ldots,\bm{\check{\mu}}_{\eta_T}) ;$\\
  $\bm{\check{\Sigma}}_\eta \gets \text{blockdiag}(\bm{\check{\Sigma}}_{\eta_1},\ldots,\bm{\check{\Sigma}}_{\eta_T});$\\
  $\check \bSigma \gets \left(\text{blockdiag}\left(\frac{1}{\sigma^2_\beta}I_q, \bm{\check{\Sigma}}_\eta\right) +{\bm C}'\bm{\check\Omega}{\bm C}\right)^{-1};$\\
    $\check\bmu \gets (\check\bmu_\beta',\check\bmu_\eta')' \gets \check\bSigma{\bm C}'{\bm\kappa};$\\
  $m \gets \sum_{t=1}^{T-1}\bm{\check{\mu}}_{\eta_{t}}'\bm{\check{\mu}}_{\eta_{t+1}}\big/\left(\sum_{t=1}^{T-1}\check{\bm\eta}_t'\check{\bm\eta}_t + \text{tr}(\check{\bSigma}_{\eta_{-T}})\right);$\\
  $\check\sigma^2_\phi\gets \check{\sigma}^2_{\eta}\big/\left(\sum_{t=1}^{T-1}\check{\bm\eta}_t'\check{\bm\eta}_t + \text{tr}(\check{\bSigma}_{\eta_{-T}})\right);$\\%
  $\ell \gets (-1 - m)/\check{\sigma}_\phi;$\\
  $u\gets (1 - m)/\check{\sigma}_\phi;$\\
  $\check\mu_\phi\gets m - \check{\sigma}_\phi\frac{\varphi(u)-\varphi(\ell)}{\Phi(u) - \Phi(\ell)} ;$\\
  $\check\mu_{\phi^2}\gets \check\mu_{\phi}^2 + \check\sigma^2_\phi\left(1-\frac{u\varphi(u)-\ell\varphi(\ell)}{\Phi(u)-\Phi(\ell)} - \left(\frac{\varphi(u)-\varphi(\ell)}{\Phi(u) - \Phi(\ell)}\right)^2\right) ;$\\
  $\check\sigma^2_{\eta_1} \gets (b + \check{\bm\mu}_{\eta_1}'\check{\bm\mu}_{\eta_1} + \text{tr}(\check{\bm\Sigma}_{\eta_1}))/(a+r/2) ;$\\ 
  $\check\sigma^2_\eta \gets (b + (1/2)(\sum_{t=2}^{T}\check{\bm\eta}_1'\check{\bm\eta}_1 +
  \text{tr}(\check{\bm\Sigma}_{\eta_{-1}} ) - 2\check{\mu}_\phi\sum_{t=1}^{T-1}\check{\bm\eta}_t\check{\bm\eta}_{t+1} +$\\
\qquad \qquad$\check{\mu}_{\phi^2}(\sum_{t=1}^{T-1}\check{\bm\eta}_t'\check{\bm\eta}_t + 
  \tr(\check{\bSigma}_{\eta_{-T}}))))/(a+r(T-1)/2);$\\
  \For{$i=1$ to $n$}{
       \For{$t=t_i,t_{i}+1, t_{i}+2$}{
      $\check{\xi}_{it} \gets (\widetilde{\bm C}_{it}'\check\bSigma {\bm C}_{it} + ({\bm C}_{it}'\check\bmu)^2)^{1/2}$}
    }
  }
\end{algorithm}
\clearpage
\bibliography{bibliography}

\begin{thebibliography}{}

\bibitem[Agresti, 2010]{agresti_2010}
Agresti, A. (2010).
\newblock {\em {Analysis of {O}rdinal {C}ategorical {D}ata}}, volume 656.
\newblock John Wiley \& Sons.

\bibitem[Albert and Chib, 1993]{albert_chib1993}
Albert, J.~H. and Chib, S. (1993).
\newblock {Bayesian {A}nalysis of {B}inary and {P}olychotomous {R}esponse
  {D}ata}.
\newblock {\em Journal of the American Statistical Association},
  88(422):669--679.

\bibitem[Albert and Chib, 2001]{albert_2001}
Albert, J.~H. and Chib, S. (2001).
\newblock {Sequential {O}rdinal {M}odeling with {A}pplications to {S}urvival
  {D}ata}.
\newblock {\em Biometrics}, 57(3):829--836.

\bibitem[Bailar, 1975]{bailar1975}
Bailar, B.~A. (1975).
\newblock The effects of rotation group bias on estimates from panel surveys.
\newblock {\em Journal of the American Statistical Association},
  70(349):23--30.

\bibitem[Bauder et~al., 2021]{bauder_2021}
Bauder, C., Luery, D., and Szelepka, S. (2021).
\newblock {Small {A}rea {E}stimation of {H}ealth {I}}nsurance {C}overage in
  2010-2019.

\bibitem[Beltr{\'a}n-S{\'a}nchez et~al., 2024]{beltran_2024}
Beltr{\'a}n-S{\'a}nchez, M.~{\'A}., Martinez-Beneito, M., and
  Corber{\'a}n-Vallet, A. (2024).
\newblock {Bayesian modeling of spatial ordinal data from health surveys}.
\newblock {\em Statistics in Medicine}.

\bibitem[Binder, 1983]{binder_1983}
Binder, D.~A. (1983).
\newblock {On the Variances of Asymptotically Normal Estimators from Complex
  Surveys}.
\newblock {\em International Statistical Review}, 51(3):279.

\bibitem[Bishop, 2006]{bishop_2006}
Bishop, C. (2006).
\newblock {\em {Pattern Recognition and Machine Learning}}.
\newblock Springer.

\bibitem[Blei et~al., 2017]{blei_2017}
Blei, D.~M., Kucukelbir, A., and McAuliffe, J.~D. (2017).
\newblock {Variational Inference: A Review for Statisticians}.
\newblock {\em Journal of the American Statistical Association},
  112(518):859--877.

\bibitem[Boes and Winkelmann, 2006]{boes_2006}
Boes, S. and Winkelmann, R. (2006).
\newblock {Ordered Response Models}.
\newblock {\em Allgemeines Statistisches Archiv}, 90:167--181.

\bibitem[Bradley et~al., 2015]{bradley_2015}
Bradley, J.~R., Holan, S.~H., and Wikle, C.~K. (2015).
\newblock {Multivariate spatio-temporal models for high-dimensional areal data
  with application to Longitudinal Employer-Household Dynamics}.
\newblock {\em The Annals of Applied Statistics}, 9(4):1761 -- 1791.

\bibitem[Brewer et~al., 1984]{brewer_early_hanif_1984}
Brewer, K., Early, L., and Hanif, M. (1984).
\newblock {Poisson, Modified {P}oisson and Collocated Sampling}.
\newblock {\em Journal of Statistical Planning and Inference}, 10(1):15--30.

\bibitem[B{\"u}rkner and Vuorre, 2019]{burkner_2019}
B{\"u}rkner, P.-C. and Vuorre, M. (2019).
\newblock {Ordinal Regression Models in Psychology: Tutorial}.
\newblock {\em Advances in Methods and Practices in Psychological Science},
  2(1):77--101.

\bibitem[Carter et~al., 2024]{carter_2024}
Carter, J.~B., Browning, C.~R., Boettner, B., Pinchak, N., and Calder, C.~A.
  (2024).
\newblock Land-use filtering for nonstationary spatial prediction of collective
  efficacy in an urban environment.
\newblock {\em The Annals of Applied Statistics}, 18(1).

\bibitem[Daly and Robinson, 2022]{daly_2022}
Daly, M. and Robinson, E. (2022).
\newblock {Depression and anxiety during COVID-19}.
\newblock {\em The Lancet}, 399(10324):518.

\bibitem[Durante and Rigon, 2019]{durante2019}
Durante, D. and Rigon, T. (2019).
\newblock {Conditionally Conjugate Mean-Field Variational Bayes for Logistic
  Models}.
\newblock {\em Statistical Science}, 34(3).

\bibitem[Fienberg, 1980]{fienberg_1980}
Fienberg, S.~E. (1980).
\newblock {The analysis of cross-classified categorical data}.
\newblock {\em Massachusetts Institute of Technology Press, Cambridge and
  London}.

\bibitem[Fienberg, 2007]{fienberg_2007}
Fienberg, S.~E. (2007).
\newblock {\em {The Analysis of Cross-classified Categorical Data}}.
\newblock Springer Science \& Business Media.

\bibitem[Gelman, 2007]{gelman2007}
Gelman, A. (2007).
\newblock {Struggles with Survey Weighting and Regression Modeling}.
\newblock {\em Statistical Science}, 22(2).

\bibitem[Gneiting and Raftery, 2007]{Gneiting_2007}
Gneiting, T. and Raftery, A.~E. (2007).
\newblock {Strictly Proper Scoring Rules, Prediction, and Estimation}.
\newblock {\em Journal of the American Statistical Association},
  102(477):359--378.

\bibitem[Hawes et~al., 2022]{hawes_2022}
Hawes, M.~T., Szenczy, A.~K., Klein, D.~N., Hajcak, G., and Nelson, B.~D.
  (2022).
\newblock Increases in depression and anxiety symptoms in adolescents and young
  adults during the {COVID-19} pandemic.
\newblock {\em Psychol. Med.}, 52(14):3222--3230.

\bibitem[Hidiroglou and You, 2016]{hid16}
Hidiroglou, M.~A. and You, Y. (2016).
\newblock {Comparison of Unit Level and Area Level Small Area Estimators}.
\newblock {\em Survey Methodology}, 42:41--61.

\bibitem[Higgs and Hoeting, 2010]{higgs_2010}
Higgs, M.~D. and Hoeting, J.~A. (2010).
\newblock A clipped latent variable model for spatially correlated ordered
  categorical data.
\newblock {\em Computational Statistics \& Data Analysis}, 54(8):1999 -- 2011.

\bibitem[Horvitz and Thompson, 1952]{Horvitz1952}
Horvitz, D.~G. and Thompson, D.~J. (1952).
\newblock {A Generalization of Sampling Without Replacement from a Finite
  Universe}.
\newblock {\em Journal of the American Statistical Association},
  47(260):663--685.

\bibitem[Hughes and Haran, 2013]{hughes2013}
Hughes, J. and Haran, M. (2013).
\newblock {Dimension reduction and alleviation of confounding for spatial
  generalized linear mixed models}.
\newblock {\em Journal of the Royal Statistical Society Series B: Statistical
  Methodology}, 75(1):139--159.

\bibitem[Jaakkola and Jordan, 2000]{jaakkola2000}
Jaakkola, T. and Jordan, M.~I. (2000).
\newblock {Bayesian Parameter Estimation via Variational Methods}.
\newblock {\em Statistics and Computing}, 10:25--37.

\bibitem[Jimenez et~al., 2023]{jimenez2023}
Jimenez, A., Balamuta, J.~J., and Culpepper, S.~A. (2023).
\newblock {A Sequential Exploratory Diagnostic Model Using a P{\'o}lya-Gamma
  Data Augmentation Strategy}.
\newblock {\em British Journal of Mathematical and Statistical Psychology},
  76(3):513--538.

\bibitem[Johnson et~al., 1994]{johnson_1994}
Johnson, N.~L., Kotz, S., and Balakrishnan, N. (1994).
\newblock {\em Continuous {U}nivariate {D}istributions}, volume~1 of {\em Wiley
  Series in Probability and Statistics}.
\newblock John Wiley \& Sons, Nashville, TN, 2 edition.

\bibitem[Kang and Kottas, 2024a]{kang_crosssectional}
Kang, J. and Kottas, A. (2024a).
\newblock {Bayesian Nonparametric Risk Assessment in Developmental Toxicity
  Studies with Ordinal Responses}.
\newblock {\em arXiv preprint arXiv:2408.11803}.

\bibitem[Kang and Kottas, 2024b]{kang_longitudinal}
Kang, J. and Kottas, A. (2024b).
\newblock {Flexible Bayesian Modeling for Longitudinal Binary and Ordinal
  Responses}.
\newblock {\em Statistics and Computing}, 34(6).

\bibitem[Kullback and Leibler, 1951]{kullback_1951}
Kullback, S. and Leibler, R.~A. (1951).
\newblock {On {I}nformation and {S}ufficiency}.
\newblock {\em The Annals of Mathematical Statistics}, 22(1):79--86.

\bibitem[Kunihama et~al., 2019]{kunihama_2019}
Kunihama, T., Halpern, C.~T., and Herring, A.~H. (2019).
\newblock {Non-parametric Bayes Models for Mixed Scale Longitudinal Surveys}.
\newblock {\em Journal of the Royal Statistical Society Series C: Applied
  Statistics}, 68(4):1091--1109.

\bibitem[Linderman et~al., 2015]{linderman2015}
Linderman, S., Johnson, M.~J., and Adams, R.~P. (2015).
\newblock {Dependent Multinomial Models Made Easy: Stick-Breaking with the
  Polya-Gamma Augmentation}.
\newblock In Cortes, C., Lawrence, N., Lee, D., Sugiyama, M., and Garnett, R.,
  editors, {\em {Advances in Neural Information Processing Systems}},
  volume~28. Curran Associates, Inc.

\bibitem[Lumley, 2004]{R_survey}
Lumley, T. (2004).
\newblock {Analysis of Complex Survey Samples}.
\newblock {\em Journal of Statistical Software}, 9(1):1--19.
\newblock R package verson 2.2.

\bibitem[Machini et~al., 2022]{Machini_2022}
Machini, B., Achia, T.~N., Chesang, J., Amboko, B., Mwaniki, P., and Kipruto,
  H. (2022).
\newblock {Cross-sectional Study to Predict Subnational Levels of Health
  Workers' Knowledge about Severe Malaria Treatment in Kenya}.
\newblock {\em BMJ Open}, 12(1):e058511.

\bibitem[McCullagh, 1980]{mccullagh_1980}
McCullagh, P. (1980).
\newblock {Regression {M}odels for {O}rdinal {D}ata}.
\newblock {\em Journal of the Royal Statistical Society: Series B
  (Methodological)}, 42(2):109--127.

\bibitem[Metin et~al., 2022]{metin_2022}
Metin, A., Erbi\c{c}er, E.~S., \c{S}en, S., and \c{C}etinkaya, A. (2022).
\newblock {Gender and {COVID-19} related fear and anxiety: {A} meta-analysis}.
\newblock {\em Journal of Affective Disorders}, 310:384--395.

\bibitem[Parker et~al., 2022]{parker_holan_janicki_2022}
Parker, P.~A., Holan, S.~H., and Janicki, R. (2022).
\newblock {{Computationally efficient Bayesian Unit-level models for
  non-Gaussian data Under informative sampling with application to estimation
  of health insurance coverage}}.
\newblock {\em The Annals of Applied Statistics}, 16(2):887 -- 904.

\bibitem[Parker et~al., 2023]{Parker_2023}
Parker, P.~A., Holan, S.~H., and Janicki, R. (2023).
\newblock {Conjugate Modeling Approaches for Small Area Estimation with
  Heteroscedastic Structure}.
\newblock {\em Journal of Survey Statistics and Methodology}.

\bibitem[Polson et~al., 2013]{polson_scott_2013}
Polson, N.~G., Scott, J.~G., and Windle, J. (2013).
\newblock {Bayesian Inference for Logistic Models Using P\'olya-Gamma Latent
  Variables}.
\newblock {\em Journal of the American Statistical Association},
  108(504):1339--1349.

\bibitem[Savitsky and Toth, 2016]{savitsky_toth_2016}
Savitsky, T.~D. and Toth, D. (2016).
\newblock {{Bayesian estimation under informative sampling}}.
\newblock {\em Electronic Journal of Statistics}, 10(1):1677 -- 1708.

\bibitem[Schliep and Hoeting, 2015]{schliep_2015}
Schliep, E.~M. and Hoeting, J.~A. (2015).
\newblock {Data augmentation and parameter expansion for independent or
  spatially correlated ordinal data}.
\newblock {\em Computational Statistics \& Data Analysis}, 90:1--14.

\bibitem[Skinner, 2018]{Skinner_2018}
Skinner, C. (2018).
\newblock {Analysis of Categorical Data for Complex Surveys}.
\newblock {\em International Statistical Review}, 87(S1).

\bibitem[Skinner, 1989]{skinner_1989}
Skinner, C.~J. (1989).
\newblock Domain means, regression and multivariate analysis.
\newblock In Skinner, C.~J., Holt, D., and Smith, T. M.~F., editors, {\em
  Analysis of Complex Surveys}, chapter~2, pages 59--84. Wiley, Chichester.

\bibitem[Sutradhar and Kovacevic, 2000]{sutradhar_2000}
Sutradhar, B.~C. and Kovacevic, M. (2000).
\newblock {Analysing ordinal longitudinal survey data: Generalised estimating
  equations approach}.
\newblock {\em Biometrika}, 87(4):837--848.

\bibitem[Thompson, 2015]{thompson2015}
Thompson, M.~E. (2015).
\newblock {Using Longitudinal Complex Survey Data}.
\newblock {\em Annual Review of Statistics and Its Application}, 2(1):305--320.

\bibitem[Tutz, 1990]{tutz_1990}
Tutz, G. (1990).
\newblock {Sequential Item Response Models with an Ordered Response}.
\newblock {\em British Journal of Mathematical and Statistical Psychology},
  43(1):39--55.

\bibitem[Tutz, 1991]{tutz_1991}
Tutz, G. (1991).
\newblock {Sequential Models in Categorical Regression}.
\newblock {\em Computational Statistics \& Data Analysis}, 11(3):275--295.

\bibitem[Tutz et~al., 2005]{tutz_2005}
Tutz, G., Simonoff, J., Kateri, M., Lesaffre, E., Loughin, T., Svensson, E.,
  Aguilera, A., Liu, I., and Agresti, A. (2005).
\newblock {The analysis of ordered categorical data: An overview and a survey
  of recent developments -- Discussion}.
\newblock {\em TEST}, 14(1):30--73.

\bibitem[Vedensky et~al., 2023]{vedensky_2023}
Vedensky, D., Parker, P.~A., and Holan, S.~H. (2023).
\newblock {Bayesian Unit-level Models for Longitudinal Survey Data under
  Informative Sampling: An Analysis of Expected Job Loss Using the Household
  Pulse Survey}.
\newblock {\em arXiv preprint arXiv:2304.07897}.

\bibitem[{World Health Organization}, 2022]{WHO_2022}
{World Health Organization} (2022).
\newblock {Mental {H}ealth and {COVID-19}: {E}arly evidence of the pandemic's
  impact}.
\newblock Technical report, World Health Organization.

\end{thebibliography}
\bibliographystyle{apalike}
\end{document}